\documentclass[%
aps,
pre,
superscriptaddress,
tightenlines,
showpacs,showkeys,
longbibliography,
notitlepage,
a4paper,12pt
]{revtex4-1}
\usepackage[english]{babel}
\usepackage{amssymb,amsmath,stmaryrd,array}

\usepackage{graphicx}
\usepackage{subfig}

\makeatletter

\newcommand{\ket}[1]{\ensuremath{|{#1}\rangle}}
\newcommand{\bra}[1]{\ensuremath{\langle{#1}|}}
\newcommand{\sca}[2]{\ensuremath{\bigl({#1}\cdot{#2}\bigr)}}
\newcommand{\avr}[1]{\ensuremath{\langle{#1}\rangle}}

\newcommand{\cnj}[1]{{#1}^{\ast}}
\newcommand{\hcnj}[1]{{#1}^{\dagger}}

\newcommand{\pdrs}[1]{\partial_{#1}}

\newcommand{\diag}{\mathop{\rm diag}\nolimits}
\newcommand{\sign}{\mathop{\rm sign}\nolimits}
\renewcommand{\Re}{\mathop{\rm Re}\nolimits}
\renewcommand{\Im}{\mathop{\rm Im}\nolimits}
\newcommand{\Tr}{\mathop{\rm Tr}\nolimits}

 \newcommand{\bs}[1]{\boldsymbol{#1}}
 \newcommand{\vc}[1]{\mathbf{#1}}
 \newcommand{\mvc}[1]{\mathbf{#1}}
 \newcommand{\uvc}[1]{\hat{\mathbf{#1}}}
 
 \newcommand{\ind}[1]{\mathrm{#1}}

\newcommand{\dd}{\mathrm{d}}

 \newcommand{\e}{\mathrm{e}}
 \newcommand{\ee}{\mathrm{e}}

\newcommand{\vac}{\mathrm{vac}}
\newcommand{\med}{\mathrm{m}}

\makeatother

\begin{document}
\DeclareGraphicsExtensions{jpg.,.eps,.png,.pdf}
\title{
  Polarization resolved radiation angular patterns of 
  orientationally ordered nanorods  
}

 \author{Alexei~D.~Kiselev}
 \email[Email address: ]{alexei.d.kiselev@gmail.com}

 \affiliation{%
  Saint Petersburg National Research University of Information Technologies,
  Mechanics and Optics (ITMO University),
  Kronverskyi Prospekt 49,
  197101 Saint Petersburg, Russia
 }

\date{\today}

\begin{abstract}
  We employ the transfer matrix approach
  combined with the Green's function method
  to theoretically study
polarization resolved far-field angular distributions of
photoluminescence from quantum nanorods (NRs) embedded in
an anisotropic polymer film.
The emission and excitation properties
of NRs are described by the emission and excitation
anisotropy tensors.
These tensors and the solution of the emission problem
expressed in terms of the evolution operators
are used to derive
the orientationally averaged coherency matrix of
the emitted wavefield.
For the case of in-plane alignment and
unpolarized excitation,
we estimate the emission anisotropy
parameter and compute
the angular profiles for the
photoluminescence polarization parameter
such as the degree of linear polarization,
the Stokes parameter $s_1$,
the ellipticity and the polarization azimuth.
We show that the alignment order parameter
has a profound effect on the angular profiles.
\end{abstract}

 \maketitle

\section{Introduction}
\label{sec:intro}

Over the last two decades
quantum nanorods (NRs)
have been the subject of intense
studies
as semiconductor nanoheterostructures
that possess a
unique combination of geometry and size dependent
emission and excitation
properties~\cite{Hu:science:2001,Shabaev:nanolett:2004,Talapin:nanolett:2003,Sitt:nanolett:2011,Gabriele:acsnano:2012,Verzzoli:acsnano:2015,Hasegawa:apl:2015,Aubert:acsphot:2015}
(see also a review~\cite{Krahne:phyrep:2011}). 
In addition to
quantum confinement effects
coming into play at length scales
comparable to the bulk exciton Bohr radius, 
these structures feature
linearly polarized photoluminescence~\cite{Hu:science:2001}
and excitation (absorption) anisotropy~\cite{Hens:jmatchem:2012,Kamal:prb:2012,Angeloni:acsph:2016}.

The linear polarization of emission
is governed by the fine structure of the ground exciton state.
It is
determined by a number of factors such as
the fine structure splittings,
the selection rules and
the exciton oscillator strengths~\cite{Efros:pra:1992,Shabaev:nanolett:2004,Talapin:nanolett:2003,Sitt:nanolett:2011,Krahne:phyrep:2011,Gabriele:acsnano:2012,Verzzoli:acsnano:2015}.
In particular,
both excitation and emission of single cadmium selenide (CdSe) quantum rods
are found to exhibit strong polarization dependence,
indicating that dipole moment exists along the long axis of the rods,
e.g., the unique $c$-axis of the wurtzite
structure~\cite{Chen:prb:2001}.

There is a variety of applications
utilizing the linear polarized emission from
the NRs that are used as efficient light emitters
for lasing~\cite{Kazes:adma:2002},
biological labeling~\cite{Yong:acsappl:2009}
and generation of nonclassical light~\cite{Pisanello:spmi:2010}.
In liquid crystal display devices,
it was found that
using NRs  as backlight source
may significantly enhance the optical efficiency of the backlighting
system~\cite{Aubert:acsphot:2015,Cunnigham:acsnano:2016,Srivastava:adma:2017}. 

Since the emission and excitation properties of
NRs crucially depend on their orientation,
it is of paramount importance
for any application utilizing the polarized emission from NRs
to control and determine their alignment in a film. 
There are several methods
to achieve unidirectional alignment
of NRs that have been
discussed in the past few
years~\cite{Neyts:optmexp:2013,Hasegawa:apl:2015,Aubert:acsphot:2015,Cunnigham:acsnano:2016}. 
One of the most promising approaches
uses the photoalignment
technique to align NRs in the liquid crystal polymer
(LCP) matrix
brought in contact with the photoaligning azo-dye layer
through the precise control over
the orientation of photosensitive dye
molecules~\cite{Du:acsnano:2015,Schneider:nanolett:2017}. 

There are several techniques developed 
for determination of the three-dimensional (3D) orientation of
the transition dipoles of single molecules.
These include polarization-sensitive detection of fluorescence
through a high-N.A. objective originally proposed
in~\cite{Fourkas:ol:2001}
and the methods based on different versions
of emission pattern imaging~\cite{Bohmer:josab:2003,Lieb:josab:2004}.

It was demonstrated that far-field polarization microscopy can
yield the 3D orientation of CdSe quantum dots~\cite{Bawendi:nat:1999}.
In Ref.~\cite{Lethiec:prx:2014} it was shown that
the 3D orientation of a single fluorescent nanoemitter can
be determined by polarization analysis of the emitted light
using the model based on the theoretical results
obtained in Refs.~\cite{Lukosz:josa:1:1977,Lukosz:josa:2:1977,Lukosz:josa:1979,Lukosz:josa:1981}.
Results of polarimetric measurements performed on
core/shell cadmium selenide/cadmium sulfide
(CdSe/CdS) dot-in-rods~\cite{Lethiec:prx:2014,Lethiec:njphys:2014}
turned out to be consistent with the hypothesis of
a linear dipole tilted with respect to the rod axis.
Orientation of gold and rare-earth-doped nanorods
was also recently studied in Refs.~\cite{Wackenhut:apl:2012,Jongwook:natnano:2017}.

Angular radiation patterns of nanoemitters
such as NRs strongly depend on its orientation.
In Ref.~\cite{Flammich:orgelectr:2010},
orientation of the emissive dipole moments
was deduced from measurements of
the far-field polarized angular radiation patterns
of organic light-emitting diodes (OLED)s
in electrical operation.
Angular distributions of polarized light from
multilayer LED structures
studied
in~\cite{Shakya:apl:2005,Schubert:apl:2007,Krames:jdt:2007,Matioli:jap:2011,Yuan:jap:2014}
are found to be important for
optimization of
light extraction efficiency and performance
of LED devices.

The radiation patterns of light emitted by NRs,
however, have received little attention
and are much less studied as compared to the LED systems.
In this work, we adapt a systematic approach and theoretically study
polarization resolved angular distributions of
photoluminescence from NRs embedded in
the liquid crystal polymer (LCP) film
and aligned by the azo-dye photoaligning layer
using the photoalignment
technique.
This geometry was previously described by
Tao Du at al. in Ref.~\cite{Du:acsnano:2015}.

One of the key features of such a multilayer system
is that both the LCP film and the azo-dye layer are 
optically anisotropic.
As it was demonstrated
for emission from hyperbolic metamaterials~\cite{Lei:srep:2014}, 
such an anisotropic environment will profoundly influence
angular radiation patterns.

Our theoretical approach
to the emission problem developed to
analyze the combined effects of
NR alignment and
optical anisotropy of surrounding media
on the angular radiation distributions
is based a  suitably modified version of the
transfer matrix method.
We show that this method
can be used in combination with the Green's function technique
to obtain our key analytical result giving
the orientationally averaged coherency matrix of NR emission
expressed in terms of the evolution operators
and the orientational averages.
The important point is that the coherency matrix
also depends on
the emission and excitation anisotropy parameters
determined by the transition dipole moments,
the level populations and the local field screening factors.
Our goal is to examine how
the angular dependence of
the polarization state of emitted light
is affected by the orientational ordering,
optical anisotropy and the emission/excitation
anisotropy parameters.

The paper is organized as follows.

In Sec.~\ref{sec:theory} we present our theoretical approach.
After introducing the emission and excitation anisotropy tensors
in Sec.~\ref{subsec:emission-excitation-tensors},
we briefly discuss the angular spectrum
representation and the evolution operators
in Sec.~\ref{subsec:evol-operator}.
Necessary details on
the transfer matrix method
are provided in Sec.~\ref{subsec:transfer-matrix}.
In Sec.~\ref{subsec:emission},
we compute the dyadic Green's function
and solve the single-emitter problem.
It is found that
the far-field eigenwave amplitudes
of the emitted wavefield can be expressed
in terms of the evolution operators.
The expression for the orientationally averaged coherency
matrix of the emitted light is obtained in
Sec.~\ref{subsec:orient-avr}.
The analytical results are employed to
perform numerical analysis of the
angular profiles for the polarization characteristics
of the emitted light in Sec.~\ref{sec:results}.
Finally, in Sec.~\ref{sec:discussion},
we draw the results together and
make some concluding remarks.
Technical details are relegated to Appendix~\ref{sec:math}.


\section{Theory}
\label{sec:theory}

\subsection{Emission and excitation anisotropy tensors}
\label{subsec:emission-excitation-tensors}

Semiconductor core/shell nanoheterostructures representing
the nanorods (NRs) studied in Ref.~\cite{Du:acsnano:2015}
are also known as the dot-in-rods 
where a spherical core is surrounded by a rod-like shell.
For the CdSe(cadmium selenide)/CdS (cadmium sulfide)
dot-in-rods with adjusted geometry of the hexagonal crystal structures,
the wurtzite $c$-axis
of both the core and the shell
is along the long axis of the rod shell,
because the growth process of the shell along the $c$-axis
is determined by the crystal anisotropy of
the core.

The band-edge exciton fine structure of CdSe nanocrystals consists of
eight states with the total angular momentum projection
on the $c$-axis $F\in\{0,\pm 1, \pm 2\}$~\cite{Efros:prb:1996,Gabriele:acsnano:2012,Verzzoli:acsnano:2015}:
$\ket{\pm 2^L}$, $\ket{\pm 1^{L,U}}$ and $\ket{0^{L,U}}$,
where the superscripts L and U indicate lower and upper sublevels,
respectively.
There are five bright (dipole allowed) exciton states:
$\ket{\pm 1^{L,U}}$ and $\ket{0^{U}}$.
For the state $\ket{0^{U}}$, the transition
dipole moment
$\bra{0}\vc{p}\ket{0^U}$,
where $\vc{p}$ is the momentum operator,
is directed along the $c$-axis (1D dipole),
whereas 
the transition
dipole moments
$\bra{0}\vc{p}\ket{\pm 1^{L,U}}$
lie in the plane normal to $\uvc{c}$ (2D dipole),
where $\uvc{c}$ is the unit vector
parallel to the $c$-axis. 

For emitted lightwave linearly polarized along
the polarization unit vector $\uvc{e}$,
the emission probabilities
for the bright exciton states
are proportional to both the magnitudes
of the projections of the transition dipoles
on the polarization vector
and the populations of the exciton states.
So, the polarization dependent
factors of the probabilities can be written in the form:
\begin{align}
  \label{eq:D-project}
  N_0 |\bra{0}\sca{\uvc{e}}{\vc{p}}\ket{0^U}|^2=D_{\parallel}\sca{\uvc{e}}{\uvc{c}}^2,
  \quad
  N_{\pm 1}|\bra{0}\sca{\uvc{e}}{\vc{p}}\ket{\pm 1^{L,U}}|^2=
  D_{\perp}^{(L,U)}\left[1-\sca{\uvc{e}}{\uvc{c}}^2\right],
\end{align}
where $\ket{0}$ stands for the ground state;
$N_{0}$ and $N_{\pm 1}$ are the level populations.
Formula~\eqref{eq:D-project} describes
the uniaxial transition anisotropy
that leads to the linearly polarized emission.
This anisotropy is governed by
a number of factors such as
the fine structure splittings, the selection rules and
the exciton oscillator strengths.
Sensitivity of these factors to the size and shape
of the nanostructures provides a way to control
the optical properties of
nanorods~\cite{Talapin:nanolett:2003,Shabaev:nanolett:2004,Gabriele:acsnano:2012,Verzzoli:acsnano:2015}. 

For nanorods embedded in surrounding
dielectric media,
both the emission and absorption
properties of NRs are
additionally influenced by
the effects of dielectric confinement
through the modification
of the interactions between charge carriers
and the local field effect~\cite{Rodina:jetp:2016}.
The latter is the effect of dielectric screening 
arising from the difference between
the external (outside) electric field and
the internal local field inside the nanostructures.
A systematic theoretical treatment of such screening
generally requires using the methods of
the effective medium theory
(details can be found, e.g., in the monographs~\cite{Choy:bk:2016,Sihlova:bk:2008}).
This theory has been applied to interpret
optical properties of
nanostructures~\cite{Hens:jmatchem:2012,Kamal:prb:2012,Gordon:josab:2014,Battie:acsph:2014,
  Angeloni:acsph:2016,Zhang:advoptm:2018,Reshetnyak:epjp:2018},
liquid crystal
systems~\cite{Kiselev:jpcm:2004,Shelestyuk:pre:2011,Fuh:lc:2013}
and hyperbolic
metamaterials~\cite{Kidwai:pra:2012,Zhang:srep:2015,Li:optica:2016,Zhang:jqsrt:2017,Lei:prb:2017}. 

In the simplest case of cylindrically symmetric and ellipsoidally shaped NRs
surrounded by an isotropic dielectric medium,
the local field effect can be described in terms of
two depolarization factors,
$N_{\parallel}$ and $N_{\perp}$,
related to the screening factors,
$L_{\parallel}=(1+N_{\parallel}(\epsilon_{\ind{em}}/\epsilon_{\med}-1))^{-1}$
and
$L_{\perp}=(1+N_{\perp}(\epsilon_{\ind{em}}/\epsilon_{\med}-1))^{-1}$,
where $\epsilon_{\med}$ ($\epsilon_{\ind{em}}$) 
 is the dielectric constant
of the surrounding medium (NR emitter),
for the electric field components
directed along and normal to the $c$-axis, respectively.
For prolate NRs with sufficiently large aspect ratio,
the component along the $c$-axis
is characterized by the smallest depolarization factor
is $N_{\parallel}$ and, in contrast to the normal components,
is almost insensitive to the screening effect.

The local field screening effect can be taken into account
by using Eq.~\eqref{eq:D-project}
with the renormalized transition coefficients:
$D_{\parallel}^{(LF)}=D_{\parallel}|L_{\parallel}|^2$
and $D_{\perp}^{(LF)}=D_{\perp}|L_{\perp}|^2$,
where $D_{\perp}=D_{\perp}^{(L)}+D_{\perp}^{(U)}$.
Clearly, this effect introduces additional
the effective transition anisotropy leading
to enhancement of the degree of linear polarization.

In a phenomenological approach,
a quantum nanoemitter (NR) is regarded
as a point oscillating dipole
located at $\vc{r}_{\ind{em}}$
with the current density
$\vc{J}_{\ind{em}}(\vc{r})=\vc{J}_{\ind{em}}\delta(\vc{r}-\vc{r}_{\ind{em}})$.
In this approach, 
the emission probability
and the above transition anisotropy
renormalized by the local field effect
can be taken into account by replacing 
the product of the current density amplitudes
with the emission dipole tensor averaged over the quantum
state of the nanoemitter.

From Eq.~\eqref{eq:D-project},
this \textit{emission anisotropy tensor} is uniaxially anisotropic
and
can be written in the following general form:
\begin{align}
  \label{eq:q-avr}
  \avr{\cnj{J_{\alpha}^{(\ind{em})}[J_{\beta}^{(\ind{em})}]}}_q\equiv J_{\alpha\beta}^{(\ind{em})}=
  J_{\perp}^{(\ind{em})}\delta_{\alpha\beta}+(J_{\parallel}^{(\ind{em})}-J_{\perp}^{(\ind{em})})c_{\alpha}c_{\beta}\equiv
  J_{\ind{em}}[\delta_{\alpha\beta}+u_{\ind{em}}c_{\alpha}c_{\beta}],
\end{align}
where $\alpha, \beta\in\{x,y,z\}$,
$\uvc{c}=(c_x,c_y,c_z)$,
$\delta_{\alpha\beta}$ is the Kronecker symbol;
an asterisk indicates complex conjugation
and
$u_{\ind{em}}=(J_{\parallel}^{(\ind{em})}-J_{\perp}^{(\ind{em})})/J_{\perp}^{(\ind{em})}$
is the \textit{emission anisotropy parameter}.
The tensor coefficients, $J_{\parallel}^{(\ind{em})}$ and
$J_{\perp}^{(\ind{em})}$,
that enter relation~\eqref{eq:q-avr}
are assumed to be
proportional to the corresponding renormalized transition coefficients
$D_{\parallel}^{(LF)}$ and $D_{\perp}^{(LF)}$.

\begin{figure*}[!tbh]
\centering
 \resizebox{90mm}{!}{\includegraphics*{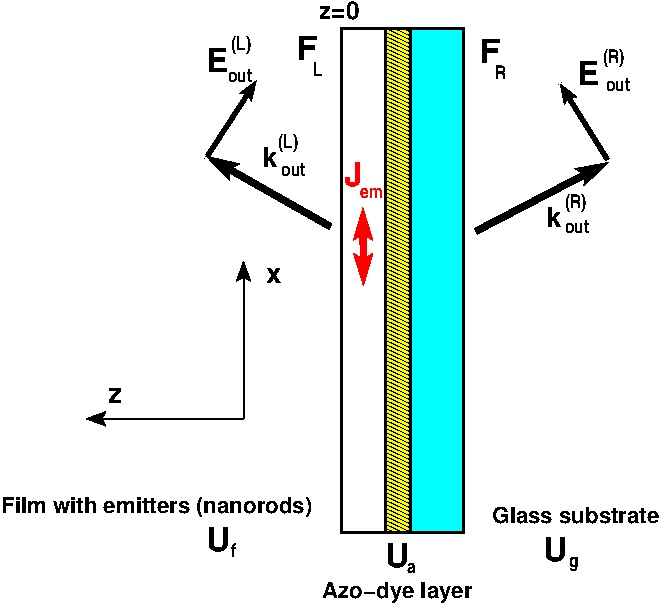}}
\caption{%
  Schematic of
emitters embedded in the top layer of the
  multi-layer structure.
}
\label{fig:emitter-geom}
\end{figure*}

The geometry of our system is
schematically depicted in Fig.~\ref{fig:emitter-geom}.
This system consists of three
layers:
(a)~the film containing the nanorods
(emitters)
and liquid crystal monomers;
(b)~the azo-dye photoaligning layer
and (c)~the glass substrate.

In order to describe 
light emission from an orientationally ordered ensemble
of nanorods
we begin with the single-emitter problem.
Solution of this problem will yield
the expressions for
the far-field eigenwave amplitudes
$E_{p}^{(\ind{em})}$ and
$E_{s}^{(\ind{em})}$.
Then the intensities of the s- and p-polarized waves registered at the detection
point are given by
\begin{align}
  \label{eq:intensities}
  I_{p,s}=J_{\ind{exc}}(\uvc{c})\avr{|E_{p,s}^{(\ind{em})}|^2}_q
\end{align}
where $J_{\ind{exc}}(\uvc{c})$ is the excitation rate
which is proportional to the
probability of absorption
and depends on the polarization of exciting light.
Similar to the emission probability,
this dependence is determined by the  excitation (absorption) anisotropy tensor
\begin{align}
  \label{eq:excitation-tensor}
  J_{\alpha\beta}^{(\ind{exc})}=J_{\perp}^{(\ind{exc})}\delta_{\alpha\beta}
  +(J_{\parallel}^{(\ind{exc})}-J_{\perp}^{(\ind{exc})})
  c_\alpha c_\beta
  \equiv
  J_{\ind{exc}}
  \left[\delta_{\alpha\beta}+u_{\ind{exc}}c_\alpha c_\beta
  \right].
\end{align}
So, for the exciting light linear polarized along the unit vector
$\uvc{e}_{\ind{exc}}$, the excitation rate is given by
\begin{align}
  \label{eq:excitation-rate}
  J_{\ind{exc}}(\uvc{c})=
  J_{\ind{exc}}
  \left[1+u_{\ind{exc}}\sca{\uvc{c}}{\uvc{e}_{\ind{exc}}}^2
  \right],
\end{align}
where
$u_{\ind{exc}}=(J_{\parallel}^{(\ind{exc})}-J_{\perp}^{(\ind{exc})})/J_{\perp}^{(\ind{exc})}$
is the \textit{excitation anisotropy parameter}.

\subsection{Operator of evolution}
\label{subsec:evol-operator}

In our subsequent calculations, 
we shall use the transfer matrix approach
which can be regarded as a modified version of 
the well-known transfer matrix method~\cite{Markos:bk:2008,Yariv:bk:2007}
and was previously applied to study the polarization-resolved conoscopic
patterns of nematic liquid crystal 
cells~\cite{Kiselev:jpcm:2007,Kiselev:pra:2008,Kiselev:jetp:2010}.
This approach has also been extended to the case of polarization
gratings and used to deduce the general expression for the effective
dielectric tensor of
deformed helix ferroelectric liquid crystal
cells~\cite{Kiselev:pre:2011,Kiselev:pre:2:2014}. 

In this approach, we deal with a harmonic electromagnetic field 
characterized by the free-space wave number
$k_{\vac}=\omega/c$,
where $\omega$ is the frequency
(time-dependent factor
is $\exp\{-i \omega t\}$),
and consider the multi-layer slab geometry
shown in Fig.~\ref{fig:emitter-geom}.
In this geometry, each optically anisotropic layer
is sandwiched between 
the bounding surfaces (substrates)
normal to the $z$ axis
and is
characterized by the dielectric tensor $\epsilon_{ij}$
(in what follows, the magnetic permittivities are equal to unity).

Further, 
we restrict ourselves to the case of stratified media
and use the angular spectrum
representation~\cite{Novotny:bk:2006,Mandl:bk:1995,Clemmow:bk:1996}
of the electromagnetic fields taken in the following form:
\begin{align}
  \label{eq:EH-form}
  \{\vc{E}(\vc{k}_P,z), \vc{H}(\vc{k}_P,z)\}=
  \int
  \{\vc{E}(\vc{r}), \vc{H}(\vc{r})\}\exp(-i \vc{k}_P\cdot\vc{r}_P)
  \dd^2\vc{r}_P
\end{align}
where
$\vc{r}=z\, \uvc{z}+\vc{r}_P$ and
the vector
\begin{align}
  \label{eq:k_p}
  \vc{k}_P/k_{\vac}=\vc{q}_P=(q_x^{(P)},q_y^{(P)},0)=
q_P(\cos\phi_P,\sin\phi_P,0)
\end{align}
represents the lateral component of the wave vector.
Then we write down
the representation for the electric and magnetic fields, $\vc{E}$ and
$\vc{H}$,
\begin{align}
  \label{eq:decomp-E}
  \vc{E}=E_z \uvc{z} +\vc{E}_{P},\quad
\vc{H}= H_z \uvc{z} +\uvc{z}\times \vc{H}_{P},
\end{align}
where the  components directed along the normal to the bounding surface
(the $z$ axis) are separated from the tangential (lateral) ones. 
In this representation,
the vectors
$\vc{E}_{P}=E_x \uvc{x}+E_y \uvc{y}\equiv
\begin{pmatrix}
  E_x\\E_y
\end{pmatrix}
$
and
$\vc{H}_{P}=\vc{H}\times\uvc{z}\equiv
\begin{pmatrix}
  H_y\\-H_x
\end{pmatrix}
$
are parallel to the substrates
and give the lateral components of the electromagnetic field.

We can now substitute the relations~\eqref{eq:decomp-E}
into the Maxwell equations
\begin{subequations}
  \label{eq:maxwell}
\begin{align}
&
  \label{eq:maxwell-E}
   \bs{\nabla}\times\vc{E}=i \mu k_{\vac} \vc{H},
\\
&
\label{eq:maxwell-H}
  \bs{\nabla}\times\vc{H}=-i k_{\vac}  \vc{D}+\frac{4\pi}{c}\vc{J}_{\ind{em}},
\end{align}
\end{subequations}
where 
$\vc{D}=\bs{\varepsilon}\cdot\vc{E}$ is the electric displacement field;
$k_{\vac}=\omega/c$ is
the \textit{free-space wave number}
and
$\vc{J}_{\ind{em}}(\vc{r})=-i\omega \bs{\mu}_{\ind{em}}\delta(\vc{r}-\vc{r}_{\ind{em}})$
is the \textit{current density of the dipole emitter}
located at $\vc{r}=\vc{r}_{\ind{em}}$,
and eliminate the $z$ components of the electric and 
magnetic fields to obtain
equations for 
the tangential components of the electromagnetic field
that can be written  
in the following $4\times 4$ matrix 
form~\cite{Kiselev:pra:2008,Kiselev:pre:2011}
(see also Appendix~\ref{sec:math}):
\begin{align}
  \label{eq:matrix-system}
  -i\pdrs{\tau}\vc{F}=\mvc{M}\cdot\vc{F}+\vc{F}_J\equiv
    \begin{pmatrix}\mvc{M}_{11}&\mvc{M}_{12}\\\mvc{M}_{21}&\mvc{M}_{22} \end{pmatrix}
    \begin{pmatrix}\vc{E}_{P}\\\vc{H}_{P} \end{pmatrix}+\vc{F}_J(\vc{k}_P)\delta(\tau-\tau_{\ind{em}}),
\quad
\tau\equiv k_{\vac} z,
\end{align}
where  
$\mvc{M}$ is the \textit{differential propagation matrix} 
 and
its $2\times 2$ block matrices
$\mvc{M}_{ij}$ are given by
\begin{subequations}
\label{eq:Mij}
   \begin{align}
&
\label{eq:Mii}
\mvc{M}^{(11)}_{\alpha\beta}=
-\epsilon_{zz}^{-1}
q_{\alpha}^{(P)}\epsilon_{z \beta},
\quad
\mvc{M}^{(22)}_{\alpha\beta}=-\epsilon_{zz}^{-1}\epsilon_{\alpha z}
q_{\beta}^{(P)},
\\
&
\label{eq:M12}
\mvc{M}^{(12)}_{\alpha\beta}=\mu 
     \delta_{\alpha \beta}-
     \frac{q_{\alpha}^{(P)}q_{\beta}^{(P)}}{\epsilon_{zz}},
\\
&
\label{eq:M21}
     \mvc{M}^{(21)}_{\alpha\beta}=\epsilon_{\alpha\beta}-
     \frac{\epsilon_{\alpha z}\epsilon_{z\beta}}{\epsilon_{zz}}
-
\mu^{-1}p_{\alpha}^{(P)}p_{\beta}^{(P)},\quad
\vc{p}_P=\uvc{z}\times\vc{q}_P.     
\end{align}
\end{subequations}

The last term on the right hand side of Eq.~\eqref{eq:matrix-system}
\begin{align}
  \label{eq:F_J}
  \vc{F}_J(\vc{k}_P)=
  \begin{pmatrix} -\vc{q}_P\epsilon_{zz}^{-1} J_z(\vc{k}_P)\\ \vc{J}_{P}(\vc{k}_P)-
\bs{\epsilon}_{z}^{\,\prime}\,
\epsilon_{zz}^{-1} J_z(\vc{k}_P) \end{pmatrix}
\end{align}
is expressed in terms of
the Fourier amplitude
of the emitter current density
\begin{align}
  &
  \label{eq:J_Fourier}
  \vc{J}(\vc{k}_P,\tau)
  =\frac{4\pi}{c}\int\vc{J}_{\ind{em}}(\vc{r}_P,z)\ee^{-i{\vc{k}_P}\cdot{\vc{r}_P}}\dd^2\vc{r}_P
    =
    \vc{J}(\vc{k}_P)\delta(\tau-\tau_{\ind{em}}),
    \quad
    \tau_{\ind{em}}\equiv k_{\vac} z_{\ind{em}},
\end{align}
where
\begin{align}
  \label{eq:J_P_ang}
  \vc{J}(\vc{k}_P)=
-4\pi i \bs{\mu}_{\ind{em}} k_{\vac}
  \ee^{-i{\vc{k}_P}\cdot{\vc{r}_{\ind{em}}}}
  =J_z(\vc{k}_P) \uvc{z}+\vc{J}_P(\vc{k}_P).
\end{align}

At $\vc{F}_J=\vc{0}$,
general solution of the homogeneous system~\eqref{eq:matrix-system}
\begin{align}
  \label{eq:evol-oprt}
  \vc{F}(\tau)=
  \mvc{U}(\tau,\tau_0)\cdot\vc{F}(\tau_0)
\end{align}
can be conveniently expressed in terms of
the \textit{evolution operator}
which is also known as the 
\textit{propagator} and is
defined as the matrix solution of 
the initial value problem
\begin{subequations}
  \label{eq:evol_problem}
\begin{align}
  \label{eq:evol_eq}
     -i\pdrs{\tau}\mvc{U}(\tau,\tau_0)
&
=
\mvc{M}(\tau)\cdot\mvc{U}(\tau,\tau_0),
\\
  \label{eq:evol_ic}
\mvc{U}(\tau_0,\tau_0)
&
=\mvc{I}_4,
\end{align}
\end{subequations}
where $\mvc{I}_n$ is the $n\times n$ identity matrix.
Basic properties of the evolution operator
are discussed in Appendix A of Ref.~\cite{Kiselev:pre:2:2014}.

For uniformly anisotropic planar structures with
the dielectric tensor of the form:
\begin{align}
  \label{eq:diel-tensor-pln}
  \epsilon_{ij}=\epsilon_z \delta_{ij} +
(\epsilon_{\parallel}-\epsilon_z)m_i m_j
+(\epsilon_{\perp}-\epsilon_z)l_i l_j,
\end{align}
where the optical axes
\begin{align}
  \label{eq:director_pln}
 \uvc{m}=(m_x,m_y,m_z)=(\cos\psi,\sin\psi,0),
\quad
\uvc{l}=\uvc{z}\times\uvc{m}=
 (-\sin\psi,\cos\psi,0)
\end{align}
lie in the plane of substrates (the $x$-$y$ plane),
the diagonal block-matrices, $\mvc{M}_{11}$ and
$\mvc{M}_{22}$, vanish,
and nondiagonal block-matrices
are given by
\begin{align}
&
  \label{eq:M12-pln}
  \mvc{M}_{12}=
  \begin{pmatrix}
    1-q_P^2/n_z^2 & 0\\
0& 1
  \end{pmatrix},
\\
&
  \label{eq:M21-pln}
  \mvc{M}_{21}=n_o^2
  \begin{pmatrix}
    1-u_a m_x^2 & -u_a m_x m_y\\
-u_a m_x m_y& 1-u_a m_y^2-q_P^2/n_o^2
  \end{pmatrix},
\end{align}
where
$\vc{q}_P=q_P\uvc{x}$;
$n_z=\sqrt{\epsilon_z}$ and
$n_o=\sqrt{\epsilon_{\perp}}$ are the principal refractive indices;
$u_a=(\epsilon_{\parallel}-\epsilon_{\perp})/\epsilon_{\perp}$
is the parameter of in-plane anisotropy.
In this case,
the operator of evolution
can be expressed in terms of 
the eigenvalue and eigenvector matrices, 
$\mvc{\Lambda}\equiv\diag(\lambda_1,\lambda_3,\lambda_3,\lambda_4)$ and $\mvc{V}$, as follows
\begin{align}
  \label{eq:U-pln}
  \mvc{U}(\tau,\tau_0)=
     \exp\{i \mvc{M}\, (\tau-\tau_0)\}=\mvc{V}
      \exp\{i \mvc{\Lambda}\, (\tau-\tau_0)\}
\mvc{V}^{-1},
\quad
\mvc{M} \mvc{V}=\mvc{V} \mvc{\Lambda},
\end{align}
where the eigenvector and eigenvalue
matrices are of the following form:
\begin{align}
&
\label{eq:V-Q-pln}
\mvc{V}=
\begin{pmatrix}
            \mvc{E} & \mvc{E}\\
             \mvc{H} & -\mvc{H}
\end{pmatrix},
\quad
\mvc{\Lambda}=\diag(\mvc{Q},-\mvc{Q}),
\quad
\mvc{Q}=\diag(q_e,q_o)
\end{align}
and can be computed
from the relations given in Appendix~B
of Ref.~\cite{Kiselev:pre:2:2014}.

\subsection{Transfer matrix}
\label{subsec:transfer-matrix}

In the ambient medium with $\epsilon_{ij}=\epsilon_{\med}\delta_{ij}$,
the general solution~\eqref{eq:evol-oprt}
can be expressed in terms of
plane waves propagating along  
the wave vectors with the tangential component~\eqref{eq:k_p}.
For such waves, the result  is given by
\begin{align}
&
  \label{eq:F_med}
  \vc{F}_{\med}(\tau)=\mvc{V}_{\med}(\vc{q}_P)
  \begin{pmatrix}
    \exp\{i \mvc{Q}_{\med}\, \tau\} & \mvc{0}\\
\mvc{0} &    \exp\{-i \mvc{Q}_{\med}\, \tau\}   
  \end{pmatrix}
  \begin{pmatrix}
    \vc{E}_{+}\\
\vc{E}_{-}
  \end{pmatrix},
\\
&
\label{eq:Q_med}
\mvc{Q}_{\med}=q_{\med}\,\mvc{I}_2,
\quad
q_{\med}=\sqrt{n_{\med}^2-q_P^2},
\end{align}
where 
$\mvc{V}_{\med}(\vc{q}_P)$
is the eigenvector matrix for the ambient medium
given by
\begin{align}
&
\label{eq:Vm-phi-q}
\mvc{V}_{\med}(\vc{q}_P)=
\mvc{T}_{\ind{rot}}(\phi_P)\mvc{V}_{\med}=
\begin{pmatrix}
 \mvc{Rt}(\phi_P)&\mvc{0}\\
\mvc{0}& \mvc{Rt}(\phi_P) 
\end{pmatrix}
\begin{pmatrix}
\mvc{E}_{\med} & -\bs{\sigma}_3 \mvc{E}_{\med}\\
\mvc{H}_{\med} & \bs{\sigma}_3 \mvc{H}_{\med}\\
\end{pmatrix},
\\
&
  \label{eq:EH-med}
  \mvc{E}_{\med}=
  \begin{pmatrix}
    q_{\med}/n_{\med}& 0\\
0 & 1
  \end{pmatrix},
\quad
 \mvc{H}_{\med}=
 \begin{pmatrix}
   n_{\med}& 0\\
0 & q_{\med}
 \end{pmatrix},
\\
&
\label{eq:Rot_matrix}
\mvc{Rt}(\phi)=\begin{pmatrix}
  \cos\phi &-\sin\phi\\
\sin\phi & \cos\phi
\end{pmatrix},
\end{align}
$\{\bs{\sigma}_1,\bs{\sigma}_2,\bs{\sigma}_3\}$ are the Pauli matrices
\begin{align}
  \label{eq:pauli}
      \bs{\sigma}_1=
      \begin{pmatrix}
        0&1\\1&0
      \end{pmatrix},
\:
      \bs{\sigma}_2=
      \begin{pmatrix}
        0&-i\\i&0
      \end{pmatrix},
\:
      \bs{\sigma}_3=
      \begin{pmatrix}
        1&0\\0&-1
      \end{pmatrix}.
\end{align}

From Eq.~\eqref{eq:F_med},
the vector  amplitudes $\vc{E}_{+}$ and
$\vc{E}_{-}$ correspond to the forward and backward eigenwaves
with
$\vc{k}_{+}=k_{\vac}(q_{\med}\,\uvc{z}+\vc{q}_P)$
and 
$\vc{k}_{-}=k_{\vac}(-q_{\med}\,\uvc{z}+\vc{q}_P)$, respectively.
Figure~\ref{fig:emitter-geom}
shows that,
in the half space $z\ge z_0$ 
after the exit face of the film
with embedded emitters
$z=z_0$,
these eigenwaves describe 
the \textit{incoming and outgoing waves}
 \begin{align}
   &
   \label{eq:in-out-plus}
   \vc{E}_{+}\vert_{z\ge z_0}=
     \vc{E}_{\ind{out}}^{(L)},
\quad
   \vc{E}_{-}\vert_{z\ge z_0}=
   \vc{E}_{\ind{in}}^{(L)}=\vc{0},
 \end{align} 
whereas,
in the half space $z\le z_3=-D$ before the entrance face of the glass substrate,
these waves are given by
\begin{align}
   \label{eq:in-out-minus}
   \vc{E}_{+}\vert_{z\le z_3}=
\vc{E}_{\ind{in}}^{(R)}=\vc{0},
\quad
   \vc{E}_{-}\vert_{z\le z_3}=
   \vc{E}_{\ind{out}}^{(R)}.
\end{align}

The boundary conditions require
the tangential components of the electric and magnetic
fields to be continuous at the boundary surfaces of  the multi-layer structure:
\begin{align}
  &
  \label{eq:continuity} 
  \vc{F}_L\equiv \vc{F}(z_0)=\vc{F}_{\med}(z_0+0)=\vc{V}_{\med}(\vc{q}_P)
  \begin{pmatrix}
    \vc{E}_{\ind{out}}^{(L)}\\
    \vc{E}_{\ind{in}}^{(L)}
  \end{pmatrix},
  \notag
  \\
  &
  \vc{F}_R\equiv \vc{F}(z_3)=\vc{F}_{\med}(z_3-0)=\vc{V}_{\med}(\vc{q}_P)
  \begin{pmatrix}
    \vc{E}_{\ind{in}}^{(R)}\\
    \vc{E}_{\ind{out}}^{(R)}
  \end{pmatrix}.
\end{align}

In the standard light scattering (transmission/reflection) problem,
we can use the boundary conditions~\eqref{eq:continuity}
to rewrite the relation
\begin{align}
  &
 \label{eq:F_LR}
\vc{F}_{L}=\mvc{U}(\tau_0,\tau_3)\cdot\vc{F}_{R},
\quad  
  \tau_i=k_{\vac} z_i,
\end{align}
in the form
\begin{align}
  \label{eq:transf-rel}
  \begin{pmatrix}
    \vc{E}_{\ind{out}}^{(L)}\\
    \vc{E}_{\ind{in}}^{(L)}
  \end{pmatrix}
=
\mvc{T}
\begin{pmatrix}
  \vc{E}_{\ind{in}}^{(R)}\\
    \vc{E}_{\ind{out}}^{(R)}
  \end{pmatrix}  
\end{align}
and introduce
the \textit{transfer matrix} $\mvc{T}$
linking the amplitudes of the eigenwaves
in the half spaces $z>z_0$ and $z<z_3$
bounded by the faces of the multilayer structure.
This matrix is the evolution operator $\vc{U}(\tau_0,\tau_3)$
in the basis of eigenmodes of the surrounding medium which
is given by
\begin{align}
  &
    \label{eq:T-op}
    \vc{T}(\tau_0,\tau_3)\equiv\vc{T}=
    \mvc{V}_{\med}^{-1}(\vc{q}_P) \mvc{U}(\tau_0,\tau_3) \mvc{V}_{\med}(\vc{q}_P)=
    \mvc{V}_{\med}^{-1} \mvc{U}_{\ind{rot}}(\tau_0,\tau_3) \mvc{V}_{\med}=
\begin{pmatrix}
\mvc{T}_{11} & \mvc{T}_{12}\\
\mvc{T}_{21} & \mvc{T}_{22}
\end{pmatrix}
,
\end{align}
where $\mvc{U}_{\ind{rot}}(\tau,\tau_0)=\mvc{T}_{\ind{rot}}(-\phi_P) \mvc{U}(\tau,\tau_0)
\mvc{T}_{\ind{rot}}(\phi_P)$
is the rotated operator of evolution.
This operator is the solution of the initial value
problem~\eqref{eq:evol_problem} 
with $\mvc{M}(\tau)$ replaced with
$\mvc{M}_{\ind{rot}}(\tau)=\mvc{T}_{\ind{rot}}(-\phi_P) \mvc{M}(\tau) \mvc{T}_{\ind{rot}}(\phi_P)$.
The scattering matrix relating the amplitudes of the incoming and
outgoing waves can be expressed in terms of the block matrices
$\vc{T}_{ij}$ giving the expressions for the transmission and
reflection matrices~\cite{Kiselev:pre:2:2014}. We shall need the
results for the case where the incident wave
$\vc{E}_{\ind{inc}}=\vc{E}_{\ind{in}}^{(L)}$ is coming from the
half-space $z>z_0$:
\begin{align}
  \label{eq:TR}
  \vc{T}_L=\vc{T}_{22}^{-1},\quad
  \vc{R}_L=\vc{T}_{12}\vc{T}_{22}^{-1},
\end{align}
where $\vc{T}_L$ and $\vc{R}_L$ are the transmission and reflection
matrices, respectively.

Our concluding remark in this section is that,
for the multi-layer structure that consists of three layers
depicted in Fig.~\ref{fig:emitter-geom},
the evolution operator and the transfer matrix
are given by
\begin{align}
  &
  \label{eq:evol-layers}
  \vc{U}(\tau_0,\tau_3)=\vc{U}_f(h_f)\vc{U}_a(h_a)\vc{U}_g(h_g),
  \quad
    \vc{T}(\tau_0,\tau_3)=\vc{T}_f(h_f)\vc{T}_a(h_a)\vc{T}_g(h_g), 
\end{align}
where
$\vc{U}_f(h_f)=\vc{U}_f(\tau_0,\tau_1)$ is the evolution operator
for the film containing the nanorods,
$D_f$ is the film thickness
($h_f=k_{\ind{vac}}D_f$);
$\vc{U}_a(h_a)=\vc{U}_a(\tau_1,\tau_2)$ is the evolution operator
for the azo-dye layer,
$D_a$ is the layer thickness
($h_a=k_{\ind{vac}}D_a$); 
$\vc{U}_g(h_g)=\vc{U}_a(\tau_2,\tau_3)$ is the evolution operator
for the glass substrate,
$D_g$ is the thickness
of the substrate
($h_g=k_{\ind{vac}}D_g$).

\subsection{Dyadic Green's function and emission problem}
\label{subsec:emission}

At $\vc{F}_J(\vc{k}_P)\ne\vc{0}$,
the solution of non-homogeneous system
\begin{align}
  \label{eq:F_D}
  \vc{F}_{\ind{em}}(\tau)=\vc{G}(\tau)\vc{F}_J(\vc{k}_P)
\end{align}
describing light radiation of the dipole emitter
is expressed in terms of the \textit{dyadic (matrix-valued) Green's function}
$\vc{G}(\tau)$ that can be found
by solving the following equation 
\begin{align}
  \label{eq:Green-func-eq}
  -i\pdrs{\tau}\vc{G}=\mvc{M}\cdot\vc{G}+\vc{I}_4\delta(\tau-\tau_{\ind{em}}).
\end{align}

For uniformly anisotropic layer
with the propagator given by
Eqs.~\eqref{eq:U-pln} and~\eqref{eq:V-Q-pln},
we can use the Fourier transform technique combined with
the residue calculus
(the poles at $q_{o,e}$ are shifted to the upper half of the complex
plane: $q_{o,e}\to q_{o,e}+i\delta_{+}$)
to
obtain the following expression for the Green's function:
\begin{align}
  &
    \label{eq:G-uni}
            \vc{G}(\tau)= i \vc{V}
  \begin{pmatrix}
    H(\tau-\tau_{\ind{em}})\ee^{i\vc{Q}(\tau-\tau_{\ind{em}})}&\vc{0}\\
    \vc{0}&-H(\tau_{\ind{em}}-\tau)\ee^{-i\vc{Q}(\tau-\tau_{\ind{em}})}
  \end{pmatrix}
            \vc{V}^{-1},
\end{align}
where
$H(\tau)=
\begin{cases}
  1,&\tau>0\\
  1/2,&\tau=0\\
  0,&\tau< 0
\end{cases}
$ is the Heaviside step function.

So, the electromagnetic field inside the film where
$\tau_1\le\tau\le\tau_0$ is given by
 \begin{align}
   \label{eq:F_f}
   \vc{F}_f(\tau)=\vc{U}_f(\tau,\tau_1)\vc{F}_0+\vc{F}_{\ind{em}}(\tau),
 \end{align}
 where the first term on the right hand side
 (general solution of the homogeneous system
 written in the form given by Eq.~\eqref{eq:evol-oprt})
 represents the waves
 reflected from (and transmitted through)
 the boundaries of the film.
 The continuity conditions
 at the film boundaries $\tau=\tau_0$ and $\tau=\tau_1$
 are
 \begin{align}
   \label{eq:BC-1_em}
   \vc{F}_f(\tau_0)=\vc{F}_L,\quad
   \vc{F}_f(\tau_1)=\vc{U}_a(h_a)\vc{U}_g(h_g)\vc{F}_R.
 \end{align}
 After eliminating $\vc{F}_0$ from Eq.~\eqref{eq:BC-1_em},
 we have
 \begin{align}
   &
   \label{eq:BC-2}
   \vc{F}_L-\vc{U}(\tau_0,\tau_3)\vc{F}_R=\vc{F}_{\ind{em}}(\tau_0)-\vc{U}_f(\tau_0,\tau_1)\vc{F}_{\ind{em}}(\tau_1)
   \notag
   \\
   &
     =[\vc{G}(\tau_0)-\vc{U}_f(\tau_0,\tau_1)\vc{G}(\tau_1)]\vc{F}_J.
 \end{align}
 For the Green's function given in Eq.~\eqref{eq:G-uni},
 the relation~\eqref{eq:BC-2} can be further simplified with
 the help of the identity
 \begin{align}
   \label{eq:G-ident}
   \vc{G}(\tau_0)-\vc{U}_f(\tau_0,\tau_1)\vc{G}(\tau_1)=i\sign(\tau_0-\tau_{\ind{em}})\vc{U}_f(\tau_0,\tau_{\ind{em}}).
 \end{align}
We can now use Eqs.~\eqref{eq:in-out-plus}--~\eqref{eq:F_LR}
to express the result in terms
 of the eigenwave amplitudes $\vc{E}_{\ind{out}}^{(L)}$
 and $\vc{E}_{\ind{out}}^{(R)}$ as follows
 \begin{align}
   &
   \label{eq:BC-3}
   \begin{pmatrix}
    \vc{E}_{\ind{out}}^{(L)}\\
    \vc{0}
  \end{pmatrix}
-
\mvc{T}
\begin{pmatrix}
  \vc{0}\\
    \vc{E}_{\ind{out}}^{(R)}
  \end{pmatrix}
   =\vc{T}_{\ind{em}}
   \begin{pmatrix}
    \vc{J}_H\\
    \vc{J}_E
  \end{pmatrix},
   \\
   &
  \label{eq:T-em}
  \vc{T}_{\ind{em}}=i\vc{T}_f(\tau_0,\tau_{\ind{em}})\vc{V}_{\ind{m}}^{-1}=\frac{i}{2\pi
     k_{\ind{vac}}q_{\ind{m}}}
  \begin{pmatrix}
\mvc{T}_{11}^{(\ind{em})} & \mvc{T}_{12}^{(\ind{em})}\\
\mvc{T}_{21}^{(\ind{em})} & \mvc{T}_{22}^{(\ind{em})}
\end{pmatrix},   
 \end{align}
 where $q_{\ind{m}}/n_{\ind{m}}$ is the $z$-component of the
 unit wave vector $\uvc{k}_{+}=\vc{k}_{+}/n_{\ind{m}}k_{\ind{vac}}$;
 $\vc{J}_H$ and $\vc{J}_E$ are given by
 \begin{align}
   \label{eq:J_EH}
   \vc{J}_H=-\frac{q_P}{\epsilon_{zz}}J_z
   \begin{pmatrix}
    1\\ 0
  \end{pmatrix},
   \quad
   \vc{J}_E=\vc{Rt}(-\phi_P)\vc{J}_P.
   \end{align}
   In our final step, we deduce the  expressions for the far-field amplitudes of the
   emitted (outgoing) waves, $\vc{E}_L^{(f-f)}$ and
   $\vc{E}_R^{(f-f)}$ from Eq.~\eqref{eq:BC-3}.
   The far-field asymptotic behavior of the amplitudes in a fixed
   direction $\uvc{r}=\vc{r}/r$ is known~\cite{Mandl:bk:1995,Novotny:bk:2006}
   to be determined by the plane wave amplitude of the angular
   spectrum representation with $\uvc{k}_{+}=\uvc{r}$.
So, the far-field amplitudes of the
   radiated waves, $\vc{E}_L^{(f-f)}$ and
   $\vc{E}_R^{(f-f)}$, are proportional
   to $\vc{E}_{\ind{out}}^{(L)}$ and $\vc{E}_{\ind{out}}^{(R)}$
   multiplied by
   the $z$-component of
   $\vc{k}_{+}=k_{\ind{vac}}q_{\ind{m}}\uvc{z}+\vc{k}_{P}$
   and $\vc{k}_{-}=-k_{\ind{vac}}q_{\ind{m}}\uvc{z}+\vc{k}_{P}$,
   respectively.
   The result reads
 \begin{align}
   &
   \label{eq:E_L_out}
   \vc{E}_{\ind{em}}=
     \begin{pmatrix}
       E_p^{(\ind{em})}\\
       E_s^{(\ind{em})}
     \end{pmatrix}
   \equiv
\vc{E}_L^{(f-f)}=
   -2\pi i k_{\ind{vac}}q_{\ind{m}}\vc{E}_{\ind{out}}^{(L)}=
   \vc{W}_H\vc{J}_H+\vc{W}_E\vc{J}_E,
   \\
   &
        \label{eq:W_H}
     \vc{W}_H=\mvc{T}_{11}^{(\ind{em})}-\vc{R}_L\mvc{T}_{21}^{(\ind{em})},
     \quad
     \vc{W}_E=\mvc{T}_{12}^{(\ind{em})}-\vc{R}_L\mvc{T}_{22}^{(\ind{em})},
   \\
   &
     \label{eq:E_R_out}
     \vc{E}_R^{(f-f)}=2\pi i k_{\ind{vac}}q_{\ind{m}}
     \vc{E}_{\ind{out}}^{(R)}=\vc{T}_L(\mvc{T}_{21}^{(\ind{em})}\vc{J}_H+\mvc{T}_{22}^{(\ind{em})}\vc{J}_E),
 \end{align}
 where the reflection and the transmission matrices,
 $\vc{R}_L$ and $\vc{T}_L$, are defined in Eq.~\eqref{eq:TR}.
 In what follows, the emitted wavefield~\eqref{eq:E_L_out}
 will be our primary concern.

\subsection{Orientational averaging}
\label{subsec:orient-avr}

We shall assume that an ensemble of aligned
NRs can be treated as a collection
of incoherently emitting and differently oriented dipoles and the total intensity
of the emitters is a sum of the intensities.
Orientation of a nanorod is specified by
the tilt and azimuthal angles, $\theta_c$ and $\phi_c$, giving the direction of
the $c$-axis:
$\uvc{c}=(\cos\theta_c\cos\phi_c,\cos\theta_c\sin\phi_c,\sin\theta_c)$
and the total intensities of the p- and s-waves can be obtained
by averaging the intensities given in Eq.~\eqref{eq:intensities}
over orientation of the $c$-axis.

More generally,
the emitted wavefield
can be described by
the coherency matrix~\cite{Mandl:bk:1995,Born:bk:1999}
\begin{align}
  \label{eq:coherency-matrix}
\mvc{M}=\avr{J_{\ind{exc}}(\uvc{c})\vc{E}_{\ind{em}}\otimes\cnj{\vc{E}}_{\ind{em}}}=
  \begin{pmatrix}
    \avr{J_{\ind{exc}}(\uvc{c})\avr{|E_p^{(\ind{em})}|^2}_q}_{\uvc{c}}&
    \avr{J_{\ind{exc}}(\uvc{c})\avr{E_p^{(\ind{em})}\cnj{[E_s^{(\ind{em})}]}}_q}_{\uvc{c}}\\
    \avr{J_{\ind{exc}}(\uvc{c})\avr{E_s^{(\ind{em})}\cnj{[E_p^{(\ind{em})}]}}_q}_{\uvc{c}}&
    \avr{J_{\ind{exc}}(\uvc{c})\avr{|E_s^{(\ind{em})}|^2}_q}_{\uvc{c}}
    \end{pmatrix},
\end{align}
where an asterisk stands for complex conjugation,
can now be calculated using the matrix relations~\eqref{eq:E_L_out}
and~\eqref{eq:W_H}. This matrix is given by
\begin{align}
  &
  \label{eq:coh-matr-gen}
  \mvc{M}=\mvc{W}_H\avr{J_{\ind{exc}}(\uvc{c})\vc{J}_H\otimes \cnj{\vc{J}}_{H}}\hcnj{\mvc{W}}_H+
  \mvc{W}_E\vc{Rt}(-\phi_P)\avr{J_{\ind{exc}}(\uvc{c})\vc{J}_P\otimes
  \cnj{\vc{J}}_{P}}\vc{Rt}(\phi_P)\hcnj{\mvc{W}}_E,
  \\
  &
    \label{eq:J_HH}
    \avr{J_{\ind{exc}}(\uvc{c})\vc{J}_H\otimes \cnj{\vc{J}}_{H}}=
    \frac{q_P^2}{\epsilon_z^2}\begin{pmatrix}
      1&0\\
      0&0
    \end{pmatrix}\avr{J_{\ind{exc}}(\uvc{c}){J}^{(\ind{em})}_{zz}}_{\uvc{c}},
  \\
  &
    \label{eq:J_PP}
    [\avr{J_{\ind{exc}}(\uvc{c})\vc{J}_P\otimes
  \cnj{\vc{J}}_{P}}]_{\alpha\beta}=\avr{J_{\ind{exc}}(\uvc{c}){J}^{(\ind{em})}_{\alpha\beta}}_{\uvc{c}},
\end{align}
where a dagger will indicate Hermitian conjugation
and $\avr{...}_{\uvc{c}}$ stands for orientational averages.

The result of orientational averaging is expressed
in terms of the two symmetric matrices:
\begin{align}
  \label{eq:averaging}
  \mvc{C}_{\alpha\beta}^{(e)}=\avr{c_\alpha c_\beta}_{\uvc{c}},
  \quad
  \mvc{C}_{\alpha\beta}^{(ex)}=\avr{c_{\ind{exc}}^2c_\alpha c_\beta}_{\uvc{c}},
\end{align}
where $c_{\ind{exc}}=\sca{\uvc{c}}{\uvc{e}_{\ind{exc}}}$
and $\uvc{e}_{\ind{exc}}$ is the polarization unit vector of the
exciting light.
Orientational ordering is described by the eigenvalues and the
eigenvectors
(the principal axes) of these matrices. Perfectly in-plane
ordering presents the important special case with
vanishing tilt angle, $\theta_c=0$.
We shall, however, consider a more general case
with nonvanishing $\theta_c$ and assume
that the angles are statistically independent
and the principal (alignment) axes are directed along the coordinate axes.
The latter implies that the matrices~\eqref{eq:averaging}
are both diagonal.
It immediately follows that the matrix~\eqref{eq:J_PP}
is also diagonal
\begin{align}
  \label{eq:JJ-vs-gamma}
  \avr{J_{\ind{exc}}(\uvc{c})\vc{J}_P\otimes
  \cnj{\vc{J}}_{P}}=\gamma_0\vc{I}_2+\gamma_3\bs{\sigma}_3,
  \quad
  \avr{J_{\ind{exc}}(\uvc{c}){J}^{(\ind{em})}_{zz}}_{\uvc{c}}=\gamma_z
\end{align}
and  the coherency matrix~\eqref{eq:coh-matr-gen} can be written in the form
of a linear combination: 
\begin{subequations}
  \label{eq:Int_avr}
\begin{align}
  &
    \label{eq:I_avr}    
  \mvc{M}=
    J_{\ind{em}}J_{\ind{exc}}[\gamma_0
  \mvc{W}_0+\gamma_3 (\cos 2\phi_P\mvc{W}_3-\sin 2\phi_P\mvc{W}_1) +\gamma_z\mvc{W}_z],
  \\
  &
    \label{eq:V_avr}
    \mvc{W}_0=\mvc{W}_E\hcnj{\mvc{W}}_E,
    \quad
    \mvc{W}_i=\mvc{W}_E\bs{\sigma}_i\hcnj{\mvc{W}}_E,
    \quad
    \mvc{W}_z=\frac{q_P^2}{\epsilon_z^2}\mvc{W}_H
    \begin{pmatrix}
      1&0\\
      0&0
    \end{pmatrix}
         \hcnj{\mvc{W}}_H,
  \\
  &
    \label{eq:gamma_0}
    \gamma_0=1+u_{\ind{exc}}\avr{c_{\ind{exc}}^2}_{\uvc{c}}+\frac{u_{\ind{em}}}{2}\left[
\avr{c_x^2+c_y^2}_{\uvc{c}}+u_{\ind{exc}}\avr{c_{\ind{exc}}^2(c_x^2+c_y^2)}_{\uvc{c}}
    \right],
  \\
  &
    \label{eq:gamma_3}
    \gamma_3=\frac{u_{\ind{em}}}{2}\left[
\avr{c_x^2-c_y^2}_{\uvc{c}}+u_{\ind{exc}}\avr{c_{\ind{exc}}^2(c_x^2-c_y^2)}_{\uvc{c}}
    \right],
  \\
  &
  \label{eq:gamma_z}
    \gamma_z=1+u_{\ind{exc}}\avr{c_{\ind{exc}}^2}_{\uvc{c}}+u_{\ind{em}}\left[
\avr{c_z^2}_{\uvc{c}}+u_{\ind{exc}}\avr{c_{\ind{exc}}^2c_z^2}_{\uvc{c}}
    \right].
\end{align}
\end{subequations}

Orientational averages that
enter the coefficients of the linear combination~\eqref{eq:I_avr}
can be expressed in terms of the orientational parameters
characterizing ordering of aligned NRs.
For in-plane ordering, these parameters
are as follows
\begin{align}
  \label{eq:p-q}
  \avr{\cos^2\phi_c}_{\uvc{c}}=\frac{1+p}{2},
  \quad
  \avr{\sin^2\phi_c}_{\uvc{c}}=\frac{1-p}{2},
  \quad
  \avr{\cos^2\phi_c\sin^2\phi_c}_{\uvc{c}}=q,
\end{align}
where $0 \le q\le \avr{\cos^2\phi_c}_{\uvc{c}}\avr{\sin^2\phi_c}_{\uvc{c}}=(1-p^2)/4$
and $-1 \le p\le 1$ is the in-plane \textit{orientational (alignment) order parameter}.
Similar results for the averages
over the tilt angle $\theta_c$ characterizing out-of-plane deviations
of the $c$-axis are given by
\begin{align}
  \label{eq:pz-qz}
  \avr{\sin^2\theta_c}_{\uvc{c}}=p_z,
\quad
  \avr{\cos^2\theta_c}_{\uvc{c}}=1-p_z,
  \quad
  \avr{\cos^2\theta_c\sin^2\theta_c}_{\uvc{c}}=q_z,
\end{align}
where $0\le q_z\le p_z(1-p_z)$ and
$0 \le p_z\le 1$ is the out-of-plane order parameter
that equals zero in the limiting case of purely in-plane ordering.

In the subsequent section, these general formulas will be used
to perform numerical analysis.

\section{Results}
\label{sec:results}

The lightfield emitted by NRs
is generally partially polarized
and the far-field angular distributions
of its polarization parameters
are determined by the coherency matrix
\begin{align}
  &
  \label{eq:I_ps}
    \mvc{M}(q_P,\phi_P)\equiv\mvc{M}(\theta,\phi)=\frac{1}{2}\bigl[
    I(\theta,\phi)\vc{I}_2+S_1(\theta,\phi)\bs{\sigma}_3+
    \notag
  \\
  &
    S_2(\theta,\phi)\bs{\sigma}_1+S_3(\theta,\phi)\bs{\sigma}_2\bigr],
  \\
  &
    \label{eq:theta-phi}
  q_P\equiv |\vc{k}_P|/k_{\ind{vac}}=n_\med\sin\theta,\quad
  \phi_P\equiv \phi,
\end{align}
where $I(\theta,\phi)=I_p(\theta,\phi)+I_s(\theta,\phi)=\Tr \mvc{M}(\theta,\phi)$ is the total intensity
and $\{S_1,S_2,S_3\}$ are the Stokes parameters,
evaluated as a function of
the emission (detection) angle,
$\theta$,
and the azimuthal angle, $\phi=\phi_P$,
(see Eq.~\eqref{eq:k_p})
that specifies orientation of the emission
plane ($\phi$ is the angle between the alignment axis and the emission
plane).
Though the polarization state
of partially polarized radiation from nanoemitters such as NRs
with the degree of polarization
\begin{align}
  \label{eq:P-deg}
  P=\sqrt{s_1^2+s_2^2+s_3^2}=\sqrt{1-\frac{4\det\mvc{M}}{(\Tr\mvc{M})^2}},\quad
  s_i\equiv S_i(\theta,\phi)/I(\theta,\phi),
\end{align}
can be completely described using the Stokes parameters,
there is a number of the technologically important parameters
widely used to characterize the anisotropy of
photoluminescence.

One of these parameters is the degree of linear
polarization
\begin{align}
  \label{eq:DOP-deg}
  DOP=\sqrt{s_1^2+s_2^2}=\frac{I_{\ind{max}}-I_{\ind{min}}}{I_{\ind{max}}+I_{\ind{min}}}
\end{align}
where $I_{\ind{max}}$ and $I_{\ind{min}}$
are the  maximum and minimum
intensities of the curve representing
the experimentally measured intensity
of light passed through the rotating polarizer
placed in the emission beam path.
Alternatively, the Stokes parameter
\begin{align}
  \label{eq:s1}
  s_1=\frac{I_{p}-I_{s}}{I_{p}+I_{s}}
\end{align}
and the related polarization ratios
\begin{align}
  \label{eq:PR}
  R_{ps}=I_p/I_s,\quad
  R_{sp}=I_s/I_p,
\end{align}
are also used as convenient
measures characterizing the anisotropy of emission 
in terms of the intensities of the p-polarized and s-polarized
waves: $I_p\equiv I_p(\theta,\phi)=\vc{M}_{11}(\theta,\phi)$
and $I_s\equiv I_s(\theta,\phi)=\vc{M}_{22}(\theta,\phi)$.
Note that the case where the degree of linear polarization is equal to
the magnitude of $s_1$,
$DOP=|s_1|$, occurs only when
the Stokes parameter
$S_2=2\Re\mvc{M}_{21}$ vanishes
and the polarization azimuth of the polarization ellipse
\begin{align}
  \label{eq:psi_p}
  \psi_p=2^{-1}\arg (s_1+i s_2)
\end{align}
differs from zero and $\pi/2$.
The ellipticity of
the polarization ellipse characterizing
the polarized part of emission
\begin{align}
  \label{eq:ellipt}
  \epsilon_{\ind{ell}}=\tan[2^{-1}\arcsin(s_3/P)]
\end{align}
is expressed in terms of
the Stokes parameter $S_3=2\Im\mvc{M}_{21}$.

Equation~\eqref{eq:I_avr} shows that each element
of the coherency matrix is a linear combination
of the three angular profiles defined by the matrices
given in Eq.~\eqref{eq:V_avr}.
In the case of unpolarized excitation
with $c_{\ind{exc}}^2$ replaced by $c_x^2+c_y^2=1-c_z^2$,
the expressions for the coefficients of the linear combination
can be obtained from the
relations~\eqref{eq:gamma_0}--~\eqref{eq:gamma_z}
in the following form:
\begin{subequations}
  \label{eq:gamma_un}
  \begin{align}
  &
    \label{eq:gamma_03z_un}
    \gamma_0=Q_1+\frac{u_{\ind{em}}}{2} Q_2,
    \quad
    \gamma_3=\frac{u_{\ind{em}}p}{2} Q_2,
  \quad
    \gamma_z=(1+u_{\ind{em}})Q_1-u_{\ind{em}}Q_2,
  \\
  &
    \label{eq:Q_12}
    Q_1=1+u_{\ind{exc}}(1-p_z),
    \quad
    Q_2=1-p_z+u_{\ind{exc}}(1-p_z-q_z).
\end{align}
\end{subequations}
An important point is that
all the above discussed polarization characteristics
depend on the two ratios of the coefficients~\eqref{eq:gamma_03z_un}
\begin{align}
    &
      \label{eq:tgamma_un}
      \tilde{\gamma}_3=\frac{\gamma_3}{\gamma_0},\quad
      \tilde{\gamma}_z=\frac{\gamma_z}{\gamma_0}
\end{align}
that can be found as the fitting parameters
when dealing with angular profiles obtained from experimental data.
Given the values of the ratios~\eqref{eq:tgamma_un}
and the emission anisotropy parameter $u_{\ind{em}}$,
we can use the relations
\begin{align}
  &
  \label{eq:Q-from-gamma}
  Q\equiv\frac{Q_2}{Q_1}=1-p_z+\frac{u_{\ind{exc}}(p_z(1-p_z)-q_z)}{1+u_{\ind{exc}}(1-p_z)}
    =\frac{2(1+u_{\ind{em}}-\tilde{\gamma}_z)}{u_{\ind{em}}(2+\tilde{\gamma}_z)},
  \\
  &
  \label{eq:p-from-gamma}
  p=\frac{(3+u_{\ind{em}})\tilde{\gamma}_3}{1+u_{\ind{em}}-\tilde{\gamma}_z}  
\end{align}
to estimate the alignment order parameters,
$p$ and $p_z$.

\begin{figure*}[!tbh]
  \centering
    \subfloat[DOP vs emission angle $\theta$ at $p=0.87$]{
   \resizebox{80mm}{!}{\includegraphics*{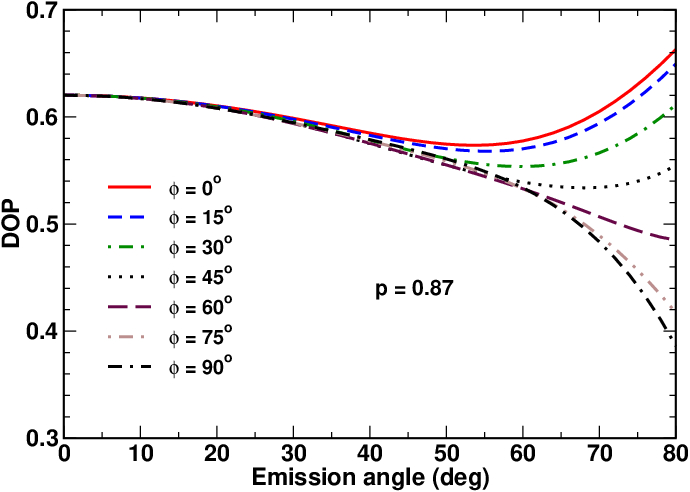}}
   \label{fig:DOP_087}
}
   \subfloat[$s_1$ vs emission angle $\theta$ at $p=0.87$]{
   \resizebox{80mm}{!}{\includegraphics*{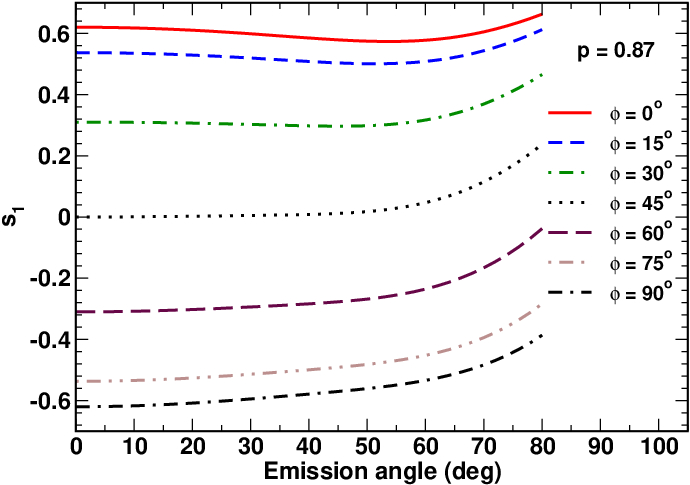}}
   \label{fig:s1_087}
}
\caption{Angular profiles of (a)~DOP and (b)~$s_1$
  computed at different values of the azimuthal angle $\phi$
  for $p=0.87$, $p_z=0$ and $u_{\ind{em}}=3.85$.
    } 
  \label{fig:DOP_s1_087}
\end{figure*}

\begin{figure*}[!tbh]
  \centering
    \subfloat[Ellipticity vs emission angle $\theta$ at $p=0.87$]{
   \resizebox{80mm}{!}{\includegraphics*{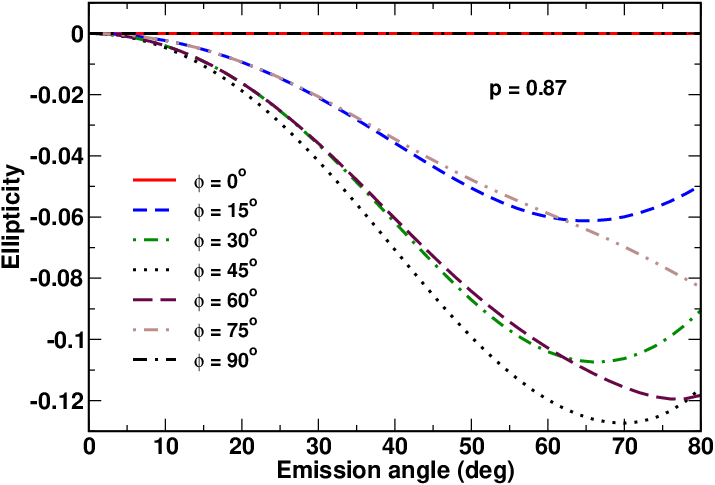}}
   \label{fig:ell_087}
}
   \subfloat[Polarization azimuth vs $\theta$ at $p=0.87$]{
   \resizebox{80mm}{!}{\includegraphics*{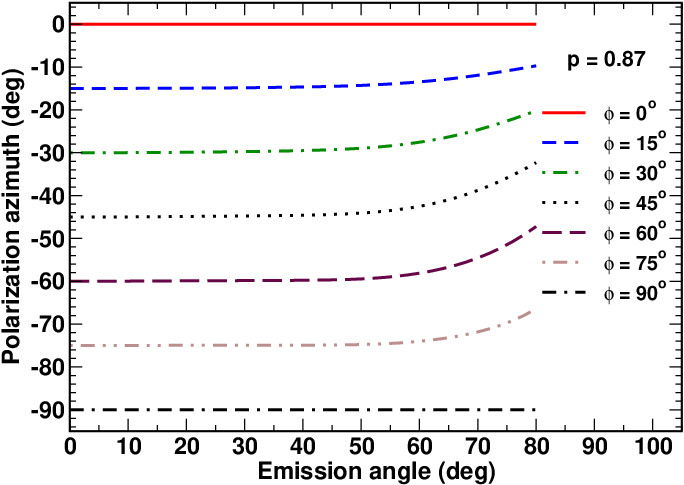}}
   \label{fig:psi_087}
}
\caption{Angular profiles of (a)~ellipticity and (b)~polarization azimuth
  computed at different values of the azimuthal angle $\phi$
  for $p=0.87$, $p_z=0$ and $u_{\ind{em}}=3.85$.
    } 
  \label{fig:ell_psi_087}
\end{figure*}

\begin{figure*}[!tbh]
  \centering
    \subfloat[DOP vs emission angle $\theta$ at $p=0.4$]{
   \resizebox{80mm}{!}{\includegraphics*{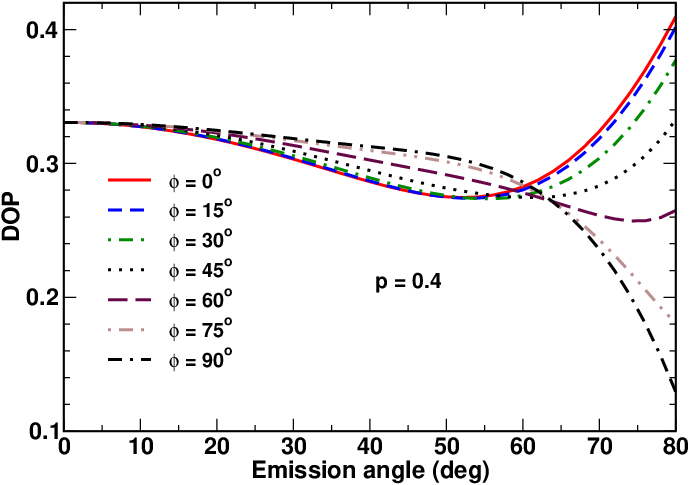}}
   \label{fig:DOP_040}
}
   \subfloat[$s_1$ vs emission angle $\theta$ at $p=0.4$]{
   \resizebox{80mm}{!}{\includegraphics*{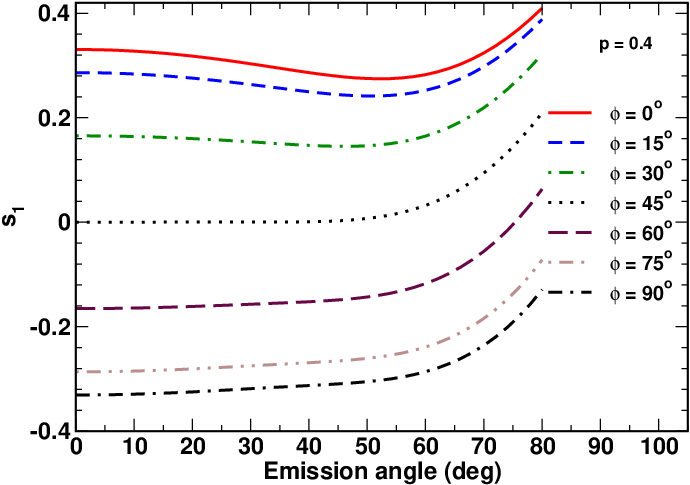}}
   \label{fig:s1_040}
}
\caption{Angular profiles of (a)~DOP and (b)~$s_1$
  computed at different values of the azimuthal angle $\phi$
  for $p=0.4$, $p_z=0$ and $u_{\ind{em}}=3.85$.
    } 
  \label{fig:DOP_s1_040}
\end{figure*}

\begin{figure*}[!tbh]
  \centering
    \subfloat[Ellipticity vs emission angle $\theta$ at $p=0.4$]{
   \resizebox{80mm}{!}{\includegraphics*{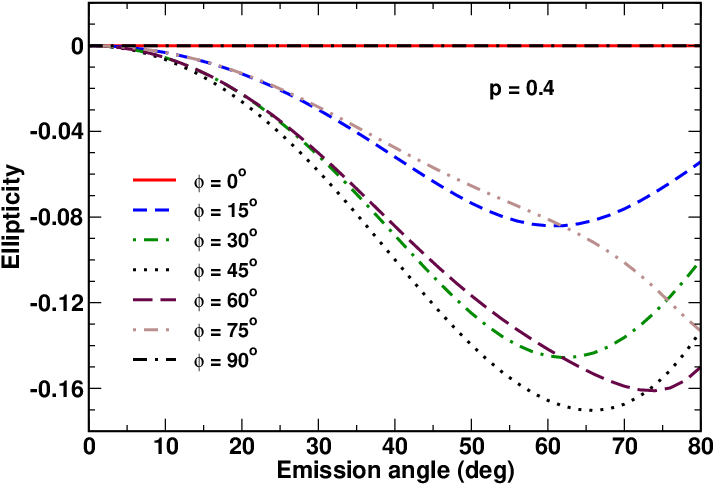}}
   \label{fig:ell_040}
}
   \subfloat[Polarization azimuth vs $\theta$ at $p=0.4$]{
   \resizebox{80mm}{!}{\includegraphics*{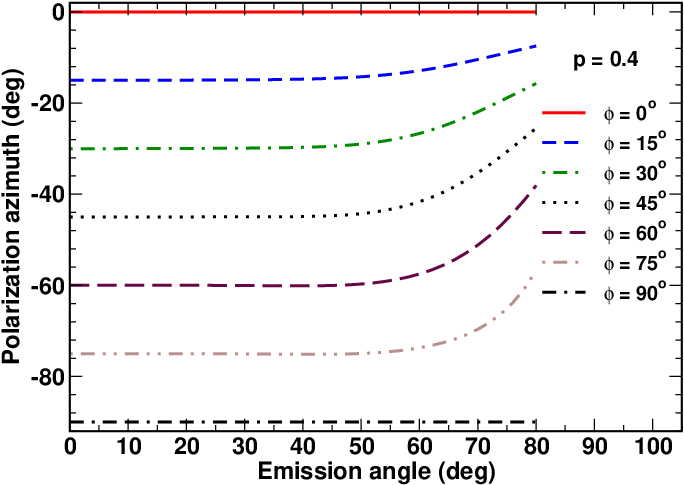}}
   \label{fig:psi_040}
}
\caption{Angular profiles of (a)~ellipticity and (b)~polarization azimuth
  computed at different values of the azimuthal angle $\phi$
  for $p=0.4$, $p_z=0$ and $u_{\ind{em}}=3.85$.
    } 
  \label{fig:ell_psi_040}
\end{figure*}

\begin{figure*}[!tbh]
  \centering
    \subfloat[DOP vs emission angle $\theta$ at $p=0$]{
   \resizebox{80mm}{!}{\includegraphics*{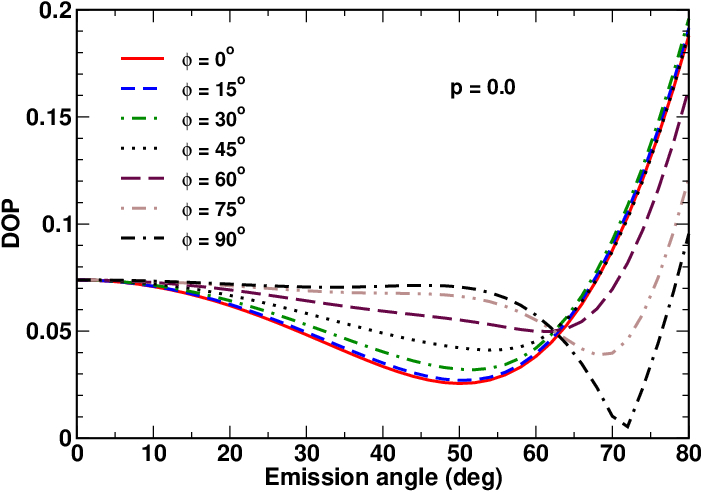}}
   \label{fig:DOP_00}
}
   \subfloat[$s_1$ vs emission angle $\theta$ at $p=0$]{
   \resizebox{80mm}{!}{\includegraphics*{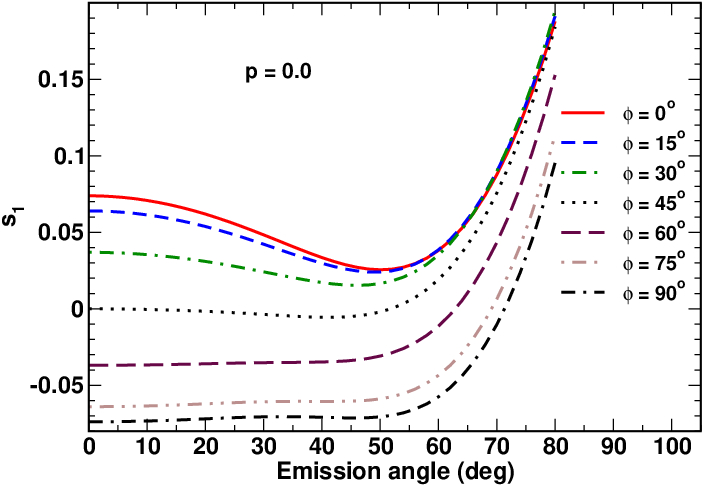}}
   \label{fig:s1_00}
}
\caption{Angular profiles of (a)~DOP and (b)~$s_1$
  computed at different values of the azimuthal angle $\phi$
  for $p=p_z=0$ and $u_{\ind{em}}=3.85$.
    } 
  \label{fig:DOP_s1_00}
\end{figure*}

\begin{figure*}[!tbh]
  \centering
    \subfloat[Ellipticity vs emission angle $\theta$ at $p=0$]{
   \resizebox{80mm}{!}{\includegraphics*{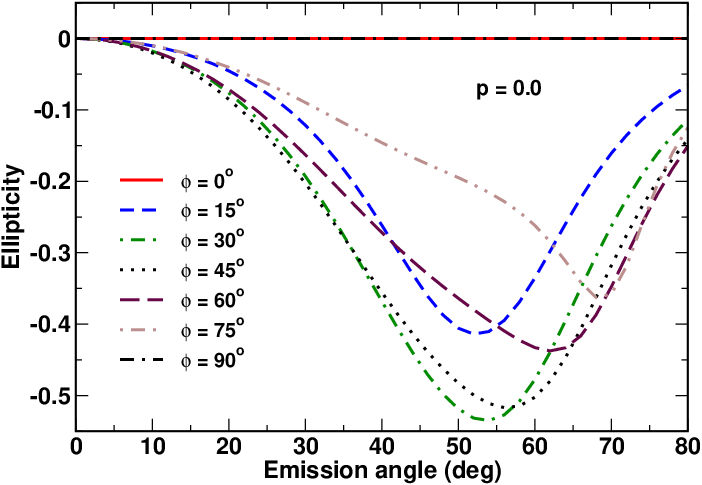}}
   \label{fig:ell_00}
}
   \subfloat[Polarization azimuth vs $\theta$ at $p=0$]{
   \resizebox{80mm}{!}{\includegraphics*{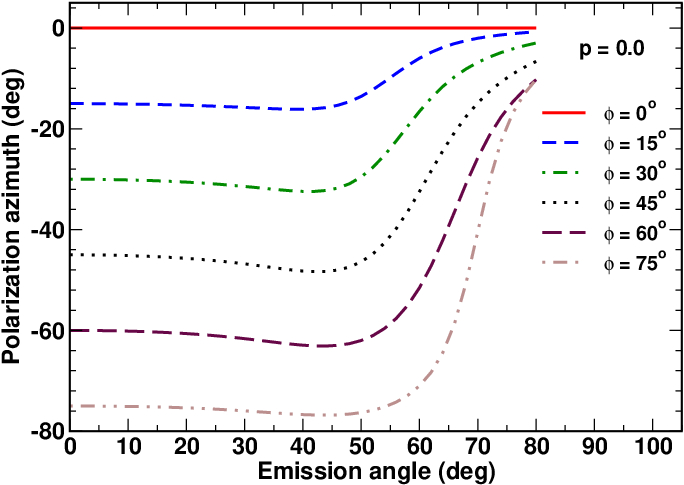}}
   \label{fig:psi_00}
}
\caption{Angular profiles of (a)~ellipticity and (b)~polarization azimuth
  computed at different values of the azimuthal angle $\phi$
  for $p=p_z=0$ and $u_{\ind{em}}=3.85$.
    } 
  \label{fig:ell_psi_00}
\end{figure*}

In what follows we concentrate on
the four polarization parameters:
the degree of linear polarization
(DOP) given by Eq.~\eqref{eq:DOP-deg},
the Stokes parameter $s_1$ (see Eq.~\eqref{eq:s1}),
the polarization azimuth, $\psi_p$,
(see Eq.~\eqref{eq:psi_p})
and the ellipticity $\epsilon_{\ind{ell}}$
(see Eq.~\eqref{eq:ellipt}).
Figures~\ref{fig:DOP_s1_087}--~\ref{fig:ell_psi_00}
present the results
for these parameters evaluated as functions
of the emission angle $\theta$
at different values of the azimuthal angle $\phi$
(the vector $(-\sin\phi,\cos\phi,0)$ is normal to the emission plane).
Calculations are performed for the emission wavelength
$\lambda_{\ind{em}}=590$~nm
assuming that
the in-plane extraordinary and ordinary refractive indexes
for the azo-dye photoaligning layer of the
thickness $D_a=15$~nm
(the data are taken from Ref.~\cite{Du:acsnano:2015})
are $n_{\parallel}^{(a)}=1.8$
and $n_{\perp}^{(a)}=1.5$
(the refractive indices are estimated from the data fitting performed in
Ref.~\cite{Kiselev:pre:2009}),
whereas
the indexes for the LCP film 
(for modeling purposes, the thickness is assumed to be $D_f=375$~nm)
containing NRs with an aspect ratio $5:1$
(the NR length is about $20$~nm and the NR diameter is about $4$~nm)~\cite{Du:acsnano:2015}
are $n_{\parallel}^{(f)}=1.7$
and $n_{\perp}^{(f)}=1.5$
(typical values for LCs), respectively.
Since the LCP molecules are aligned along
the easy axis of the SD1 layer, which is perpendicular
to the alignment axis of NRs
(we assume perfectly in-plane ordering of NRs with $p_z=0$
and the alignment axis directed along the $x$ axis)~\cite{Du:acsnano:2015},
the latter is normal to the in-plane optic axes
of both the SD1 layer and the LCP film
so that their orientation
with respect to the emission plane is defined by
the azimuthal angle $\psi$
(see Eq.~\eqref{eq:director_pln})
which is equal to $\pi/2-\phi$.

Figures~\ref{fig:DOP_s1_087} and~\ref{fig:ell_psi_087}
show the angular profiles computed
for highly ordered NRs embedded into
the LCP film using the value of
the NR alignment order parameter $p=0.87$
reported in Ref.~\cite{Du:acsnano:2015}.
In addition,
the value of DOP measured in~\cite{Du:acsnano:2015}
at $\theta=\phi=0$ is about $0.62$.
From this result, we have obtained the estimate
for the emission anisotropy parameter:
$u_{\ind{em}}\approx 3.85$
giving the value of $u_{\ind{em}}$ used in our calculations.

Referring to Fig.~\ref{fig:DOP_087},
when the azimuthal angle $\phi$ does not exceed $\pi/4$,
DOP is generally a nonmonotonic function of the emission
angle
that increases at sufficiently large values of $\theta$.
At $\phi>\pi/4$,
DOP monotonically decreases with $\theta$.

In the special case where $\theta=0$ and
the detected lightwave is propagating along the normal to the substrates,
the value of DOP is independent of $\phi$.
As it can be seen from Fig.~\ref{fig:ell_087},
at $\theta=0$, the polarized part of the emitted light is
linearly polarized and the ellipticity vanishes.
In this case, rotation of the emission plane about the $z$ axis
by the azimuthal angle $\phi$
is equivalent to the rotation of the sample by the angle
$-\phi$. As a result, the polarization plane is rotated by the same
angle and, as is indicated in Fig.~\ref{fig:ell_psi_087},
the polarization azimuth $\psi_p$ equals $-\phi$.

Another consequence of such rotation is that,
at $\theta=0$,
the Stokes parameter $s_1$
changes from $DOP(0)=0.62$ to $-DOP(0)=-0.62$
as $\phi$ varies from zero to $\pi/2$.
It immediately follows from the relation
\begin{align}
  \label{eq:DOP-s1}
  s_1+i s_2=DOP\e^{2 i\psi_p}
\end{align}
that defines the DOP and the polarization azimuth
given by Eqs.~\eqref{eq:DOP-deg}
and~\eqref{eq:psi_p}, respectively.
This effect can be seen from the curves presented in
Fig.~\ref{fig:s1_087}.
This figure also demonstrates that, in agreement with the
identity~\eqref{eq:DOP-s1},
$s_1(0)$ becomes negative when $\phi>\pi/4$.
In this region, $s_1$ is a monotonically increasing function of
$\theta$, whereas both the magnitude of $s_1$, $|s_1|$,
and the DOP fall as the emission angle increases.

Equation~\eqref{eq:DOP-s1} shows that
the magnitude of $s_1$ is generally smaller than
the DOP and the difference between the DOP and $s_1$
is dictated by the polarization azimuth $\psi_p$.
The curves presented in Fig.~\ref{fig:psi_087}
illustrate a noticeable increase in the polarization azimuth as
$\theta$ becomes sufficiently large.

The angular profiles for the ellipticity
of the polarized part of the emission
plotted in Fig.~\ref{fig:ell_087}
are computed from Eq.~\eqref{eq:ellipt}.
It is found that, for the cases where
the alignment axis is either parallel or normal to the emission
plane ($\phi=0$ and $\phi=\pi/2$, respectively),
the light appears to be linearly polarized and the ellipticity
equals zero.
At $0<\phi<\pi/2$,
this is, however, no longer the case
and the ellipticity shows generally nonmonotonic variations with
$\theta$. The ellipticity magnitude
$|\epsilon_{\ind{ell}}|$ reaches its highest value at $\phi=\pi/4$.
For $p=0.87$, this value is about $0.12$.

In Figs.~\ref{fig:DOP_s1_040} and~\ref{fig:ell_psi_040},
we show what happen when the alignment order parameter
is reduced and present the results computed at $p=0.4$.
An important point is that the most
part of the above discussion
being independent of the alignment order parameter
remains valid for the case of poorly
aligned NRs.
So, we will focus our attention on
the differences introduced by changes in
orientational ordering of NRs.

Referring to Fig.~\ref{fig:DOP_040},
reduction of the order parameter has a detrimental effect on the DOP.
In particular, the value of $DOP(0)$ is reduced to $0.33$.
As is seen from Fig.~\ref{fig:ell_040},
this effect also manifests itself in an increase
of the largest value of the ellipticity magnitude
which is now about $0.17$.
Qualitatively,
the common feature shared by all the angular profiles calculated at $p=0.4$
is that degree of variations and
nonmonotic behavior become more pronounced
as compared to the case with $p=0.87$.

Figures~\ref{fig:DOP_s1_00} and~\ref{fig:ell_psi_00}
demonstrate that
this feature is further enhanced in the limiting case
of completely disordered NRs with $p=0$.
As is shown in Fig.~\ref{fig:DOP_s1_00},
at $p=0$
the values of both the DOP and $|s_1|$
experience significant reduction.
For instance, $DOP(0)$ is about $0.07$.
By contrast, the largest value of the ellipticity magnitude
(see Fig.~\ref{fig:ell_00}) is increased up to $0.51$.
Angular profiles for the polarization azimuth
plotted in Fig.~\ref{fig:psi_00} demonstrate strong variations
and rapidly approach the vicinity of zero
as the emission angle increases.
Clearly, these features can be regarded as
the effects of optically anisotropic environment.

Our concluding remark in this section
concerns the parity symmetry
of the polarization parameters under the change
of sign of the emission and azimuthal angles:
$\theta\to -\theta$ and $\phi\to -\phi$.
All the polarization parameters are even functions of $\theta$:
$DOP(-\theta,\phi)=DOP(\theta,\phi)$,
$s_1(-\theta,\phi)=s_1(\theta,\phi)$,
$\epsilon_{\ind{ell}}(-\theta,\phi)=\epsilon_{\ind{ell}}(\theta,\phi)$
and $\psi_p(-\theta,\phi)=\psi(\theta,\phi)$.
Similarly, when $\phi$ changes its sign,
$DOP$ and $s_1$ remain intact:
$DOP(\theta,-\phi)=DOP(\theta,\phi)$ and
$s_1(\theta,-\phi)=s_1(\theta,\phi)$.
In contrast, the ellipticity and the polarization azimuth
are even functions of $\phi$:
$\epsilon_{\ind{ell}}(\theta,-\phi)=-\epsilon_{\ind{ell}}(\theta,\phi)$
and $\psi_p(\theta,-\phi)=-\psi_p(\theta,\phi)$.

\section{Discussion and conclusions}
\label{sec:discussion}

In this paper,
we have studied the far-field angular
distributions of the polarization parameters
characterizing the anisotropy of photoluminescence
from orientationally ordered NRs placed inside
an optically anisotropic multilayer system.
By using a suitably modified version of
the transfer matrix method combined
with the Green's function technique
we have found the solution of the emission problem
expressed in terms of the evolution operators
(see Eqs.~\eqref{eq:BC-3}--~\eqref{eq:W_H}).
The emission and excitation
anisotropy tensors
(see Eq.~\eqref{eq:q-avr} and Eq.~\eqref{eq:excitation-rate},
respectively) that depend
on the transition dipole moments,
the exciton level populations and the local field screening factors
are then used to
deduce the expression for
the  orientationally averaged coherency matrix~\eqref{eq:coherency-matrix}
of the emitted optical field.

The results of our theoretical analysis
are applied to the geometry of the photoalignment method where
aligned NRs are embedded into the LCP film
placed on top of the photoaligning azo-dye layer~\cite{Du:acsnano:2015}.
By assuming perfectly in-plane ordering of nanorods and unpolarized excitation,
we have computed the degree of linear polarization (DOP)
(see Eq.~\eqref{eq:DOP-deg}),
the Stokes parameter $s_1$ (see Eq.~\eqref{eq:s1}),
the polarization azimuth (see Eq.~\eqref{eq:psi_p})
and the ellipticity (see Eq.~\eqref{eq:ellipt}) of the emitted light
as functions of the emission and azimuthal angles,
$\theta$ and $\phi$,
(see Eq.~\eqref{eq:theta-phi})
at different values of the alignment order parameter $p$
(see Eq.~\eqref{eq:p-q}).
We have found that the values of DOP and the order parameter
measured in Ref.~\cite{Du:acsnano:2015} can be
used to obtain the estimate for the emission anisotropy parameter $u_{\ind{em}}$:
$u_{\ind{em}}\approx 3.75$.

Angular profiles computed as
$\theta$ dependencies
of the polarization parameters in differently oriented  emission
planes at varying value of $p$
are presented in Figs.~\ref{fig:DOP_s1_087}--~\ref{fig:ell_psi_00}.
These profiles are shown to be determined by
the two coefficients given by Eq.~\eqref{eq:tgamma_un}
that depend on the NR orientational averages and the
emission/excitation anisotropy parameters, while
the components of the three matrices given
by Eq.~\eqref{eq:V_avr}
define the angular dependent contributions
governed by the optical anisotropy
of the LCP film and the photoaligning layer.

We have evaluated the profiles 
for the three cases:
(a)~the film with highly ordered NRs
($p=0.87$);
(b)~the film with poorly ordered NRs
($p=0.4$);
and
(c)~the film with disordered NRs
($p=0.0$).
It is found that reduction of orientational order
has a detrimental effect on the DOP and the Stokes parameter
$s_1$ so that their values for the disordered NRs
are an order of magnitude smaller than for the highly ordered NRs.
By contrast, the largest value of ellipticity significantly grows
as the alignment order parameter decreases.
The curves computed at $p=0$
(see Figs.~\ref{fig:DOP_s1_00} and~\ref{fig:ell_psi_00})
demonstrate a pronounced nonmonotonic behavior
that can be regarded as the effect of
the optically anisotropic environment.

It should be stressed that the emission and
absorption properties of NRs
embedded in surrounding dielectric media
are generally influenced by
the effects of dielectric confinement~\cite{Rodina:jetp:2016}.
In our phenomenological approach,
NRs are treated as radiating dipoles and
these properties are 
described by the emission and excitation
anisotropy tensors.
The dielectric confinement effects
for dot-in-rods placed in optically anisotropic media
has not been studied in any detail yet
and analysis of such effects
is well beyond the scope of this paper.

We conclude this paper with the remark
on the local field effect in the anisotropic LCP film.
As it was discussed in Sec.~\ref{sec:theory}
this effect will typically enhance
the anisotropy of emission and excitation
leading to nonzero anisotropy parameters
$u_{\ind{em}}$ and $u_{\ind{exc}}$
even if the transition anisotropy is vanishing.
For NRs with an aspect ratio $5:1$ and
$\epsilon_{\ind{em}}\equiv\epsilon_{CdS}=5.23$~\cite{Ninomiya:jap:1:1995}
embedded in the LCP film
with $\epsilon_{\parallel}^{(f)}=2.62$
and $n_{\perp}^{(f)}=2.16$,
the screening factors can be estimated
assuming that the $c$-axis ($\uvc{c}=\uvc{x}$)
is normal to the optic axis of the film
directed along the $y$-axis.
To this end we can use the well-known
analytical results for an optically isotropic ellipsoid
placed in the anisotropic medium~\cite{Sihlova:bk:2008}
and evaluate the components,
$L_x$, $L_y$ and $L_z$
of the screening factor tensor (dyadic). 
In our case, this tensor is biaxial.
Its diagonal elements are: $L_z\approx 0.61<L_y\approx 0.67<L_x\approx 0.93$
giving the anisotropy ratios: $(L_x/L_z)^2\approx 2.33$,
$(L_x/L_y)^2\approx 1.94$
and $(L_z/L_y)^2\approx 0.83$.

These results show that the anisotropic
environment plays the role of a symmetry breaking factor
and, as a consequence, the transition anisotropy tensor~\eqref{eq:q-avr}
is no longer uniaxial.
For instance, in the case of in-plane alignment with $p_z=0$,
the generalized form of this tensor is as follows 
\begin{align}
  \label{eq:J_biaxial}
  \vc{J}_{\ind{em}}=J_{\ind{em}}(\vc{I}+u_{\ind{em}}\uvc{c}\otimes\uvc{c}+u_{z}\uvc{z}\otimes\uvc{z}),
\end{align}
where $\vc{I}$ is the identity matrix.
The effect of the local field giving the small negative value of
$u_z=(L_z/L_y)^2-1$ might be further enhanced
by the transition anisotropy factor.
This may have a profound effect on the angular profiles
of the polarization parameters of radiated field.
Since
the effects of the anisotropic
dielectric confinement on the optical properties of
nanostructures have not been the subject of intense studies,
the physics behind such a phenomenological approach
is poorly understood and requires a more sophisticated
theoretical treatment.

\begin{acknowledgments}
  This work was partially supported by
  the Russian Science Foundation under grant 19-42-06302.
\end{acknowledgments}

\appendix

\section{Equations for lateral components}
\label{sec:math}

In this section we discuss how to exclude
the $z$-components of the electromagnetic field,
$E_z$ and $H_z$,
that enter the representation~\eqref{eq:decomp-E},
from the Maxwell equations~\eqref{eq:maxwell}.
Our task is  to derive the closed system of equations
for the lateral (tangential) components,
$\vc{E}_P$ and $\vc{H}_P$.

After substituting Eq.~\eqref{eq:decomp-E}
into the Maxwell equations~\eqref{eq:maxwell},
we use
decomposition for the differential operator that enter
Eq.~\eqref{eq:maxwell}
\begin{align}
  \label{eq:decomp-nabla}
  k_{\vac}^{-1} \bs{\nabla}=\uvc{z}\,\pdrs{\tau} +i \bs{\nabla}_{P},
\quad
\bs{\nabla}_{P}^{\perp}= \uvc{z}\times \bs{\nabla}_{P},
\end{align}
where $\tau=k_{\vac} z$;
$
\bs{\nabla}_{P}=-i\, k_{\vac}^{-1}\,
(\uvc{x}\,\pdrs{x}+\uvc{y}\,\pdrs{y})\equiv
(\nabla_x,\nabla_y)
$
and
$
\bs{\nabla}_{P}^{\perp}=
(\nabla_x^{\perp},\nabla_y^{\perp})=(-\nabla_y,\nabla_x),
$
to recast
Maxwell's equations~\eqref{eq:maxwell}
into the following form:
\begin{subequations}
  \label{eq:maxwell-p-1}
\begin{align}
&
\label{eq:mxwll-Ep-1}
-i\pdrs{\tau}\,
   [\uvc{z}\times\vc{E}_{P}]=\mu  \vc{H}
-
\bs{\nabla}_{P}\times\vc{E},
\\
&
\label{eq:mxwll-Hp-1}
-i\pdrs{\tau}\,
   \vc{H}_{P}=\vc{D}
+
     \bs{\nabla}_{P}\times\vc{H}+{\vc{J}},
     \quad
     {\vc{J}}\equiv\frac{4\pi i}{c k_{\vac}}\vc{J}_D,
\end{align}
\end{subequations}
where the explicit expressions for the last terms on the right hand side of the
system~\eqref{eq:maxwell-p-1}
are as follows
\begin{subequations}
  \label{eq:aux-np}
\begin{align}
&
  \label{eq:aux-np-E}
\bs{\nabla}_{P}\times\vc{E}
=
-\bs{\nabla}_{P}^{\perp} E_z+
\sca{\bs{\nabla}_{P}^{\perp}}{\vc{E}_{P}}\,
\uvc{z},
\\
&
\label{eq:aux-np-H}
\bs{\nabla}_{P}\times\vc{H}
=
-\bs{\nabla}_{P}^{\perp} H_z+
\sca{\bs{\nabla}_{P}}{\vc{H}_{P}}\,
\uvc{z}.
\end{align}
\end{subequations}

We can now substitute 
the electric displacement field 
and the current density of the dipole
$\vc{J}=-4\pi i \bs{\mu}_D\delta(\vc{r}-\vc{r}_0)$
written as a sum of
the normal and in-plane components
\begin{align}
\label{eq:decomp-D}
  \vc{D}=D_z \uvc{z} +\vc{D}_{P},
\quad
\vc{J}=J_z \uvc{z} +\vc{J}_{P},  
\end{align}
into Eq.~\eqref{eq:mxwll-Hp-1}
and derive the following expression for
its $z$-component
\begin{align}
  &
\label{eq:D_z}
D_z=
\epsilon_{zz} E_z+
\sca{\bs{\epsilon}_{z}}{\vc{E}_{P}}
=
-\sca{\bs{\nabla}_{P}}{\vc{H}_{P}}-{J}_z,
\end{align}
where $\bs{\epsilon}_{z}=(\epsilon_{zx},\epsilon_{zy})$.

From Eqs.~\eqref{eq:maxwell-p-1}
and~\eqref{eq:D_z},
it is not difficult to deduce the relations
\begin{align}
&
   \label{eq:H_z}
H_z=
\mu^{-1}
\sca{\bs{\nabla}_{P}^{\perp}}{\vc{E}_{P}}
\\
&
\label{eq:E_z}
 E_z
=
-\epsilon_{zz}^{-1}
\bigl[
\sca{\bs{\epsilon}_{z}}{\vc{E}_{P}}
+\sca{\bs{\nabla}_{P}}{\vc{H}_{P}}
+{J}_z
\bigr]
\end{align}
linking the normal (along the $z$ axis) and  the lateral
(perpendicular to the $z$ axis) components. 

By using the relation~\eqref{eq:E_z},
we obtain
the tangential component of the field~\eqref{eq:decomp-D}
\begin{align}
&
\label{eq:D_p}
\vc{D}_{P}=
\bs{\epsilon}_{z}^{\,\prime} E_z
+
\bs{\varepsilon}_z\cdot \vc{E}_{P}
=
\bs{\varepsilon}_{P}\cdot \vc{E}_{P}
-
\bs{\epsilon}_{z}^{\,\prime}\,
\epsilon_{zz}^{-1}
\bigl[
\sca{\bs{\nabla}_{P}}{\vc{H}_{P}}
+{J}_z
\bigr]
\end{align}
where
$
\bs{\varepsilon}_z=
\begin{pmatrix}
  \epsilon_{xx}& \epsilon_{xy}\\
\epsilon_{yx} & \epsilon_{yy} 
\end{pmatrix}$;
$\bs{\epsilon}_{z}^{\,\prime}=(\epsilon_{xz},\epsilon_{yz})$
and the effective dielectric tensor, $\bs{\varepsilon}_{P}$, 
for the lateral components is given by
\begin{align}
  \label{eq:diel-p}
 \bs{\varepsilon}_{P}
=
 \bs{\varepsilon}_{z}-
\epsilon_{zz}^{-1}\,
\bs{\epsilon}_{z}^{\,\prime}\otimes
\bs{\epsilon}_{z}.
\end{align}

Maxwell's equations~\eqref{eq:maxwell-p-1} 
can now be combined with the relations~\eqref{eq:aux-np}
to yield the system
\begin{subequations}
  \label{eq:maxwell-p-2a}
\begin{align}
&
\label{eq:mxwll-Ep-2a}
-i\pdrs{\tau}\,
   \vc{E}_{P}= \mu  \vc{H}_{P}
+ \bs{\nabla}_{P} E_z,
\\
&
\label{eq:mxwll-Hp-2a}
-i\pdrs{\tau}\,
   \vc{H}_{P}=\vc{D}_P
-
\bs{\nabla}_{P}^{\perp} H_z+{\vc{J}}_P,
\end{align}
\end{subequations}
where $H_z$, $E_z$
and $\vc{D}_P$ are given in
Eq.~\eqref{eq:H_z},
Eq.~\eqref{eq:E_z} and
Eq.~\eqref{eq:D_p}, respectively.

So, this system immediately gives the final result
\begin{subequations}
  \label{eq:maxwell-p-2}
\begin{align}
&
  \label{eq:mxwll-Ep-2}
-i\pdrs{\tau}\,
   \vc{E}_{P}=
-
\bs{\nabla}_{P}
[\epsilon_{zz}^{-1}\sca{\bs{\epsilon}_{z}}{\vc{E}_{P}}]
+
\mu  \vc{H}_{P}
-
\bs{\nabla}_{P}
                \epsilon_{zz}^{-1}
                [\sca{\bs{\nabla}_{P}}{\vc{H}_{P}}+{J}_z],
\\
&
\label{eq:mxwll-Hp-2}
-i\pdrs{\tau}\,
   \vc{H}_{P}=
\bs{\varepsilon}_{P}\cdot \vc{E}_{P}
-
\bs{\nabla}_{P}^{\perp}
[\sca{\bs{\nabla}_{P}^{\perp}}{\vc{E}_{P}}/\mu]
-
\bs{\epsilon}_{z}^{\,\prime}\,
     \epsilon_{zz}^{-1}
     [\sca{\bs{\nabla}_{P}}{\vc{H}_{P}}+{J}_z]+{\vc{J}}_P,
\end{align}
\end{subequations}
that can be easily rewritten in
the matrix form used in Sec.~\ref{sec:theory}.


\end{document}